\documentclass[a4paper,11pt]{article}
\pdfoutput=1 

\usepackage{jcappub} 

\usepackage[T1]{fontenc} 
\usepackage{epsf}
\usepackage{graphicx}  
\usepackage{dcolumn}   
\usepackage{bm}        
\usepackage{mathrsfs}
\usepackage{hyperref,graphicx,amsfonts,amssymb,amsthm,amsmath,psfrag}
\usepackage{subfigure}
\usepackage{comment}
\usepackage{enumerate}
\usepackage{booktabs}
\usepackage{tikz}
\usepackage{siunitx}
\usepackage{float}
\usepackage[normalem]{ulem}	
\newcommand\redout{\bgroup\markoverwith
{\textcolor{red}{\rule[.5ex]{8pt}{0.8pt}}}\ULon}
\usepackage{xcolor}
\usepackage{color}
\usepackage{todonotes}

\usetikzlibrary{decorations.markings,snakes}
\tikzset{
  fermionline/.style={line width=1pt,postaction={decorate},
    decoration={markings,
      mark=at position 0.5 with {\draw[-stealth] (0,0)--(2pt,0);}}},
  bosonline/.style={line width=1pt,decorate,

    decoration={snake,amplitude=1,segment length=4}},
  higgsline/.style={line width=1pt,dashed}
}
\tikzset{
particle/.style={thin,draw=blue, postaction={decorate},
decoration={markings,mark=at position .5 with {\arrow[blue]{stealth}}}},
gluon/.style={decorate, draw=black, decoration={snake=coil}}
}

\def \be {\begin{equation}}
\def \ee {\end{equation}}
\def \bi {\begin{enumerate}}
\def \ei {\end{enumerate}}

\title{\boldmath Combined Preheating on the lattice with applications to Higgs inflation}

\author[a]{Jo\"{e}l Repond,}
\author[a,b]{Javier Rubio}

\affiliation[a]{Institut de Th\'{e}orie des Ph\'{e}nom\`{e}nes Physiques, 
\'{E}cole Polytechnique F\'{e}d\'{e}rale de Lausanne, CH-1015 Lausanne, 
Switzerland}
\affiliation[b]{Institut f\"ur Theoretische Physik, Ruprecht-Karls-Universit\"at Heidelberg,
Philosophenweg 16, 69120 Heidelberg, Germany.}

\emailAdd{joel.repond@cern.ch}
\emailAdd{j.rubio@thphys.uni-heidelberg.de}

\abstract{We use classical lattice simulations in 3+1 dimensions to study the interplay between the resonant 
production of particles during preheating and the subsequent decay of these into a set of secondary species. We choose to work in a simplified version of Higgs inflation in which the Higgs field non-minimally coupled to gravity plays the role of the inflaton. Our numerical results extend the analytical estimates in the literature beyond the linear regime and shed some light on the limitations of the analytical techniques. The inclusion of fast and inefficient decays postpones the onset of parametric resonance by depleting 
the particles produced at the bottom of the potential. In spite of this delay, fermions are shown to play an important role on the destruction of the inflaton field. The limitations of our approach and its applications to a realistic Higgs inflation scenario are also discussed.}

\begin{document}
\maketitle
\flushbottom
\section{Introduction}

Our description of the early Universe is based on two important pillars: inflation and big bang nucleosynthesis. In spite of the good understanding of these two periods, the connection between them is weakly constrained by observations. 

At the end of inflation, most of the energy of the Universe is stored in the zero mode of the inflaton field. All other fields have been diluted by the exponential expansion and are homogeneous up to small quantum fluctuations.  In order for the thermal history of the Universe to start, the energy sitting in the inflaton condensate must be transferred to the Standard Model (SM) particles in a direct or indirect way. The relocation of energy can proceed via perturbative decays~\cite{Albrecht:1982mp,Dolgov:1982th,Abbott:1982hn} or involve highly non-perturbative effects such as parametric resonance \cite{Dolgov:1990abcd,Brandenberger1990abcd,Kofman:1994rk,Brandenberger1995abcd,Kofman:1997yn,Greene:1997fu} or tachyonic preheating~\cite{Felder:2000hj,Felder:2001kt}.
Depending on the model under consideration, these mechanisms can appear in isolation or coexist. 

A non-trivial interplay between resonant particles production and perturbative decays is likely to happen in all models of inflation in which the inflaton field is coupled to some bosonic species which are themselves coupled to fermions in a non-negligible way~\cite{Kasuya:1996np,GarciaBellido:2008ab,Bezrukov:2008ut,Mukaida:2012bz}. The oscillations of the inflaton around the minimum of the potential give rise to time dependent masses for all the particles coupled to it and allow for particle creation. While the creation of fermions directly out of the inflaton condensate is severely restricted by Pauli blocking effects, the occupation numbers of bosons can grow without limit. The production mechanism is indeed an \textit{induced} mechanism: the number of bosons created at each oscillation of the inflaton field is proportional to the number of particles previously existing. If the created particles are stable, this property translates 
into an exponential growth of the population. The situation changes completely if the created bosons are not
completely stable, but can rather decay into some secondary fermionic species. The decay into fermions will tend to deplete the boson occupation numbers, suppressing with it the Bose enhancement~\cite{Kasuya:1996np,GarciaBellido:2008ab,Bezrukov:2008ut,Mukaida:2012bz}. The energy exchange between the different components will depend on the value and hierarchy of the couplings. On general grounds, we can distinguish three types of models:
 \begin{enumerate}[i)]
 \item \textit{Models with slow decays}: If the decay rate of bosons into fermions is small compared with
 the oscillation period of the inflaton field, the depletion of bosons will be negligible and the development 
 of parametric resonance will proceed as if the fermions  were absent. This is what happens for instance if  one
 considers the decay of the Higgs field in models where the inflationary dynamics is driven
 by some physics beyond the SM \cite{Enqvist:2013kaa,Enqvist:2014tta,Figueroa:2015rqa,Enqvist:2015sua}.
 
\item \textit{Models with fast and efficient decays}: If both the decay rate \textit{and} the fraction of energy per oscillation going into fermions are large, reheating could take place before the full development of parametric resonance, like in instant preheating~\cite{Felder:1998vq}. A particular model within this category is MSSM inflation \cite{Allahverdi:2011aj}, where roughly the 20 $\%$ of the energy density of the condensate is drained out of it in every oscillation of the inflaton field. 
 
\item \textit{Models with fast and inefficient decays}: If the primary bosons created out of the inflaton condensate are significantly depleted \textit{but} the fraction of energy going into fermions is small, perturbative and non-perturbative effects will be inevitably mixed. 
The reheating of the Universe will have to wait for the amplitude of the inflaton to be small enough as to allow the primary products to accumulate. A paradigmatic example of this \textit{Combined Preheating} scenario is Higgs inflation, where the Higgs field non-minimally coupled to gravity plays the role of the inflaton. An analytical formalism for \textit{Combined Preheating}  was presented in  Refs.~\cite{GarciaBellido:2008ab,Bezrukov:2008ut}. This formalism can be applied to the first stages of preheating, where the backreaction  and rescatterings of the created quanta into the inflaton condensate can be neglected. Once the number of particles becomes large, the coherence of the oscillations is destroyed and fully non-linear simulations are required in order to follow the evolution of the system.   

\end{enumerate}
In this work, we consider a numerical implementation of the \textit{Combined Preheating} idea on a 3+1 expanding lattice. Among the different models that might be considered, we choose to work in the context of Higgs inflation.
A rigorous analysis of preheating in this model would require the implementation of non-Abelian gauge interactions with the Standard Model structure and a proper treatment of quarks and leptons on the lattice. Since the main purpose of this paper is not obtaining a realistic output for Higgs inflation, but rather the lattice characterization of \textit{Combined Preheating}, we will overpass the above difficulties by considering a simplified version of the Higgs inflation model: 
\begin{enumerate}
\item We will replace the Higgs-gauge boson interactions arising from the SM covariant derivatives by global couplings among the Higgs field and three scalar degrees of freedom playing the role of the SM gauge bosons. 
We will assume that these three scalars are non mutually- nor self- interacting. 
\item We will not implement fermions in an explicit way. The boson decay into quarks and leptons will be emulated by introducing friction terms proportional to the gauge boson decay widths in the corresponding equations of motion. In order to ensure energy conservation, we will assume that the decay products are  ultrarelativistic and will derive a consistent evolution equation for the associated energy density.  
\end{enumerate}

This paper is organized as follows. In Section \ref{sec1} we review the Higgs inflation model and 
present the particular set of interactions to be considered in our numerical simulations. The implementation 
and output of these simulations is presented in Section \ref{sec2}, with Section \ref{sec2_1} and \ref{sec2_2} 
devoted respectively to the non-perturbative production of bosons in the absence and presence of fermions. The
robustness of our simulations versus lattice artifacts is considered in Section \ref{sec2_3}. Section~\ref{sec3} 
discusses the limitations of our simplified treatment and the applicability of the results to the realistic Higgs 
inflation scenario. The conclusions are presented in Section \ref{sec_conc}. 

 \section{Higgs inflation: a case of study}\label{sec1}
 
Higgs inflation is based on a minimalistic extension of the SM Lagrangian ${\cal L}_{SM}$ which includes a non-minimal coupling between the Higgs field and gravity, namely  \cite{Bezrukov:2007ep}
\be
\label{lagr0}
S_{HI}=\int  d^4x\sqrt{-g}\,{\cal L}_{HI}     \hspace{10mm}\textrm{with}\hspace{10mm}
{\cal L}_{HI}={\cal L}_{SM}+ \left(\frac{M_P^2}{2}+\xi H^\dagger H\right)  R \,.
\ee
Here $H$ stands for the Higgs doublet, $R$ is  the Ricci scalar, $M_P=1/\sqrt{8 \pi G}=2.435 \times 10^{18}$~GeV is the reduced Planck mass and the non-minimal coupling $\xi$ is assumed to be in the range $1 \ll \xi\ll M_P^2/v_{EW}^2$ with $v_{EW}$ the vacuum expectation value of the Higgs field. 

\subsection{Inflationary phase}\label{sec1_1}
The  relevant part for inflation in \eqref{lagr0} is the scalar-gravity sector, which, in the unitary gauge $H=(0, h/\sqrt{2})^T$, takes the form
\begin{equation}\label{SGsector}
S=\int d^4x\sqrt{-g}\,
 \left[\frac{M_P^2+\xi h^2}{2}  R-\frac{1}{2}(\partial h)^2-\frac{\lambda}{4}\left(h^2-v_{EW}^2\right)^2\right] \,.
\end{equation}
The fact that \eqref{SGsector} can give rise to inflation becomes apparent when the theory is written in the so-called Einstein frame. This frame is obtained by performing a conformal transformation $g_{\mu\nu}\rightarrow \tilde g_{\mu\nu}=\Omega^2(h) g_{\mu\nu}$ with conformal factor  $\Omega^2(h)= 1+\xi h^2/M_P^2$. The procedure gives rise to a non-minimal kinetic term for the Higgs field, which can be recast into a canonical form by considering the field redefinition 
 \begin{equation}\label{confR}
 \frac{d\chi}{d h}=\sqrt{\frac{\Omega^2+6\xi^2h^2/M_P^2}{\Omega^4}}\hspace{10mm}  \underset{\xi\gg 1}\longrightarrow \hspace{10mm} 
\chi\simeq \begin{cases} 
 h\,, & h< \frac{M_P}{\xi}\,,\\
\sqrt{\frac{3}{2}} M_P \log \Omega^2(h)\,, & h >\frac{M_P}{\xi}\,. 
   \end{cases}
    \end{equation}
Taking into account the previous expressions we can rewrite the Einstein-frame version of the Higgs symmetry breaking potential  
\begin{equation}\label{potE}
V(h(\chi))=\frac{1}{\Omega^4[h(\chi)]}\frac{\lambda}{4}\left(h^2(\chi)-v_{EW}^2\right)^2\,,
\end{equation}
as\footnote{Note that the vacuum expectation value of the Higgs field can be safely neglected during the inflationary and preheating stages considered in this paper. The hierarchy between the electroweak and the Planck scales in addition to the restriction $\xi\ll M_P^2/v_{EW}^2$ allow us to approximate $1+\xi v_{EW}^2/M_P^2\approx 1$, which  significantly simplifies the expression of the potential at $\chi\gg M_P/\xi$.}
 \begin{equation}\label{potentialE1}
 V(\chi)\simeq
\begin{cases} 
      \frac{\lambda}{4}(\chi^2-v_{EW}^2)^2\,, & \chi<\frac{M_P}{\xi}\,, \\
\frac{\lambda M_P^4}{4\xi^2}\left(1-e^{-\sqrt{2/3}\,\vert\chi\vert /M_P}\right)^2\,, & \chi> \frac{M_P}{\xi}\,.
   \end{cases}
 \end{equation}
The flatness of the potential at large field values allows for inflation with the standard chaotic initial conditions.  The normalization of the spectrum of primordial scalar perturbations $V/\epsilon=24\pi^2\Delta_{\cal R}^2 M_P^4 \simeq (0.0276\, M_P)^4$ fixes the relation $\xi\simeq 47000 \sqrt{\lambda}$. The non-minimal coupling $\xi$ can be therefore determined from observations once the value of the Higgs self-coupling at the scale of inflation is specified. 

At tree level, the model predicts a spectral tilt $n_s\simeq 0.97$ and a small tensor to scalar ratio $r\simeq 0.0034$, in excellent agreement with the latest CMB results~\cite{Ade:2015lrj,Array:2015xqh}. These predictions are universal and hold even in the presence of quantum corrections if i) the renormalization prescription preserves the shift symmetry of the Einstein frame 
potential at large field values\footnote{This asymptotic symmetry is the Einstein frame manifestation of 
the approximate scale symmetry of Eq.~\eqref{lagr0} at $h \gg M_P/\sqrt{\xi}$. A procedure for removing the divergencies arising from radiative corrections while keeping scale symmetry intact at all orders in perturbation theory was presented in Refs.~\cite{Shaposhnikov:2008xi,Armillis:2013wya,Gretsch:2013ooa} (see also \cite{Herrero-Valea:2016jzz}). The embedding 
of Higgs inflation in this type of scale-invariant framework and its self-consistency was considered in
Refs.~\cite{GarciaBellido:2011de,Bezrukov:2012hx,GarciaBellido:2012zu,Rubio:2014wta,Trashorras:2016azl}. For results in other prescriptions not satisfying this requirement cf. Refs.~\cite{Barvinsky:2008ia, DeSimone:2008ei,Barvinsky:2009fy,Barvinsky:2009ii}.}~\cite{Bezrukov:2009db,Bezrukov:2010jz} and ii) the Higgs self-coupling is larger than a critical value at the inflationary scale~\cite{Bezrukov:2014bra} (see Ref.~\cite{Bezrukov:2014ipa} for a review).

\subsection{Preheating phase} \label{sec1_2}
Inflation ends when the value of the Higgs field $\chi$ drops below the Planck scale. After this point, the Higgs starts to oscillate around the minimum of the potential which, for field values $M_P/\xi<\chi <\sqrt{3/2} M_P$, can be well approximated by the quadratic expression\begin{equation}\label{quadraticP}
V(\chi)\simeq \frac{1}{2}M^2\chi^2\,,\hspace{10mm} \textrm{with} \hspace{10mm}  M =\sqrt{\frac{\lambda}{3}} \frac{M_P}{\xi}\,.
\end{equation} 
 The relation between the Higgs field and the SM particles is dictated by the SM symmetry group and the non-minimal coupling to gravity. As shown in Refs.~\cite{GarciaBellido:2008ab,Bezrukov:2008ut}, the transfer of energy from the inflaton to the SM particles proceeds via a \textit{Combined Preheating} stage involving both resonant particle production and perturbative decays. 
 
 The main mechanism draining energy from the inflaton condensate is the non-perturbative creation of SM gauge bosons\footnote{The occupation number of fermions is severely restricted by Pauli-blocking effects. The associated  distribution cannot experience parametric resonance and the fermions are forced to evolve towards higher momenta~\cite{Greene:1998nh,Giudice:1999fb,GarciaBellido:2000dc,Greene:2000ew,Peloso:2000hy,Berges:2010zv}. On the contrary, the occupation numbers of  bosons can grow without limit due to bosonic amplification~\cite{Kofman:1994rk,Kofman:1997yn,Greene:1997fu}.} at the bottom of the inflationary potential.  Particle production takes place whenever the adiabaticity condition $\vert \dot{\tilde m}_{B_i}\vert\ll \tilde m_{B_i}^2$ is violated~\cite{Kofman:1997yn,GarciaBellido:2008ab,Bezrukov:2008ut}. The quantity
\begin{equation}\label{masses}
\tilde{m}^2_{B_i} (t)  \equiv \frac{{m_{B_i}}^2}{\Omega^2}=\frac{g_i^2 M_P^2(1-e^{-\sqrt{2/3}\vert\chi(t)\vert/M_P})}{4\xi}
\end{equation}
is the Einstein-frame version of the SM gauge bosons masses $m_{B_i}=g_i h/2$. Here we have adopted a compact notation in which $g_i=g_2,g_2$ and $g_2/\cos\theta_W$ for the $B_i=W^+,W^-$ and $Z$ bosons respectively, $\theta_W=\tan^{-1}(g_1/g_2)$ is the weak mixing angle and $g_1$ and $g_2$ are the gauge couplings associated to the Standard Model  $U(1)_Y$ and $SU(2)_L$ gauge groups. 

\noindent Once produced the gauge bosons tend to decay into the SM fermions they are coupled to at rates proportional to their effective masses \eqref{masses}
\begin{equation}\label{decayWZ}
\Gamma_{B_i} (\chi)=\alpha_{B_i} \tilde{m}_{B_i}(\chi)\,,
\end{equation}
with $\alpha_{B_i}$ a constant. The large expectation value of the Higgs field at the end of inflation ($\chi\sim M_P$) makes this decay very efficient. This translates into an important depletion of the gauge bosons created during the first oscillations that prevents the onset of parametric resonance. In the absence of this enhancement, the fraction of energy transferred to the SM particles is very small. Note however that the amplitude of the Higgs field decreases with time due to the expansion of the Universe and to particle creation. This diminishes the depletion power of perturbative decays and opens the possibility of developing parametric resonance at later times: the smaller the rate, the larger the number of gauge bosons that survive the decay and the larger the fraction of energy that can be taken out of the Higgs condensate via bosonic stimulation. Whether the fermions will dominate over the Higgs condensate before or after the onset of parametric resonance will depend on the precise value of the couplings at the relevant energy scale. 

\section{Numerical simulations}\label{sec2}

We will model the Higgs-gauge boson interactions arising from the SM covariant derivatives as scalar interactions between the Higgs field and three scalar components\footnote{In a clear abuse of language, we will continue to refer to these scalars as \textit{gauge bosons}.} $B_i$ playing the role of the $W^{\pm}$ and $Z$ bosons. The equivalent of non-Abelian interactions among these components will be ignored. To emulate the decay into fermions we will proceed as follows:
\begin{enumerate}[i)]
\item We will include a friction term proportional to the decay rates \eqref{decayWZ} in the equations of motion for the gauge fields. 
\item We will assume that the fermions produced in the decay are always relativistic and have an effective equation of state $p_F=\frac{1}{3}\rho_F$. 
\item We will use the previous assumption and the conservation law $\dot \rho_T+3H\left(\rho_T+p_T\right)=0$ for the total energy density and pressure of the system
\begin{equation}
\rho_T=\rho_\chi+\rho_F+\sum_{i=1}^3\rho_{B_i}\,,\hspace{10mm} \textrm{and}\hspace{10mm} p_T=p_\chi+p_F+\sum_{i=1}^3 p _{B_i}\,,
\end{equation} 
to derive an equation of motion for the fermionic energy density $\rho_F$.
\end{enumerate} 
The above procedure gives rise to a set of equations for the Higgs field, the three scalars components playing the role of the $W^+$, $W^-$ and $Z$ bosons and the energy density of fermions 
\begin{eqnarray}
&& \ddot\chi+3H\dot \chi+\left(V+\sum_{i=1}^3\frac12\tilde m_{B_i}^2B_i^2\right)_{,\chi}=\frac{1}{a^2}\nabla^2\chi\,,\label{chieq}\\
&&\ddot B_i+\left(3H+\Gamma_{B_i}\right)\dot B_i+\tilde{m}^2_{B_i} B_i  =\frac{1}{a^2}\nabla^2 B_i\,, \label{Beq}\\ 
&& \dot\rho_F + 4H\rho_F-\sum_{i=1}^3 \Gamma_{B_i}\dot B_i^2 =- \frac{1}{a^2}\left(\nabla(\dot \chi\nabla \chi)+\sum_{i=1}^3 \nabla(\dot B_i\nabla B_i)\right) \,, \label{Feq} 
\end{eqnarray}
with $V=V(\chi)$ the Higgs potential \eqref{potentialE1} at $\chi>M_P/\xi$, 
$\tilde{m}^2_{B_i}=\tilde{m}^2_{B_i}(\chi)$ the $B_i$ masses  \eqref{masses}  and 
$\Gamma_{B_i}=\Gamma_{B_i}(\chi)$ the decay rates \eqref{decayWZ}. Note that the gradient terms $\nabla(\dot\chi\nabla\chi)$ and $\nabla(\dot B_i\nabla B_i)$ on the right-hand side of  Eq.~\eqref{Feq} are full derivative terms which should not affect the value of the lattice average quantities we will be interested in.\footnote{After verifying this is indeed the case, we decided to neglect those terms in all the long-time/high-resolution simulations presented in this paper. For details, cf. Section~\ref{sec2_3}.}

The equations \eqref{chieq}-\eqref{Feq} will be implemented in a modified version of the well-known LATTICEEASY code \cite{Felder:2000hq}. The standard leapfrog method used by LATTICEEASY has interesting properties like time reversibility, symplecticity and long time stability. Unfortunately for velocity dependent forces like those associated to the friction terms in Eqs.~\eqref{Beq} and \eqref{Feq}, this algorithm becomes quickly unstable. For this reason, we decided to use an explicit Runge-Kutta scheme of order four.\footnote{We checked that our modified integrator gives the same results as the standard LATTICEEASY leapfrog integrator when the friction terms are absent.}

The value of the Higgs field and its time derivative at the onset of the reheating stage will be determined by numerically solving the homogeneous equation of motion for the Higgs field with initial slow-roll conditions. The time $t=0$ in our simulations is identified with the end of inflation and the scale factor $a=a(t)$ evolves self-consistently with the Friedmann equations\footnote{We neglect (Planck suppressed) metric fluctuations.}
\begin{equation}\label{FRW}
\frac{\ddot a}{a}=-\frac{4\pi G}{3}\langle \rho_T+3p_T\rangle\,,   \hspace{10mm}   \left(\frac{\dot a}{a}\right)^2=\frac{8\pi G}{3}\langle \rho_T\rangle\,,
\end{equation}
with $\langle \cdot \rangle$ denoting the average of the corresponding quantity over the whole simulation volume. The system of equations \eqref{chieq}-\eqref{FRW} is indeed redundant. We will make use of this redundancy as a check of energy conservation. In particular, we will evolve the scale factor with the first equation in \eqref{FRW} and will check if the second one is satisfied at any time step.

The code uses a cubic lattice of size $L$ with $N^3$ points uniformly distributed, grid spacing $dx=L/N$ and periodic boundary conditions $\phi_{Njk} = \phi_{0jk}$, $\phi_{iNk}=\phi_{i0k}$ and $\phi_{ijN} = \phi_{ij0}$, with $\phi_{ijk}$ the fields value at the lattice point $(i,j,k)$. Each temporal step involves the computation of the equations of motion at each point of the lattice and the evaluation of relevant quantities as the total energy density. In order to speed up the simulations, we will use the MPI version of LATTICEEASY with the slab decomposition \cite{Felder:2000hq}. The original version will be modified to compute 
the Laplacian with a fourth order discretization that uses the two nearest points on the lattice in each direction.

\noindent Lattice momenta are grouped in concentric shells called bins. The minimal and maximal momenta that can be resolved on the lattice are given by
\be
k_{\text{min}}=\frac{k_{\text{max}}}{N_{\text{bins}}}=\frac{2\pi}{L}\hspace{10mm}\textrm{with}\quad\quad N_{\text{bins}} = \frac{N\sqrt{3}}{2}+1\,.
\ee
We will choose the lattice parameters $L$ and $N$ in such a way that all the momenta involved are sufficiently covered during the whole simulation time. The robustness of our results with respect to modifications of these parameters will be considered in Section \ref{sec2_3}.

At the end of inflation, most of the energy of the system is stored in the zero mode of the inflaton. All other fields have been diluted by the exponential expansion and are homogeneous, up to small quantum fluctuations  \cite{Polarski:1995jg}. In order to mimic these fluctuations, we will add random initial perturbations for all the bosonic\footnote{In our simplified model, fermions are not directly coupled to the inflaton field. Since these particles appear only as secondary products of the $B_i$ bosons, we decided not to include any fluctuations on top of their initial energy density ($\rho_F=0$). By consistency of the equations, the fermions will catch up with the bosons in a few steps of the scheme.} species below a given cutoff $k_{\rm \Lambda}$ in momentum space \cite{Khlebnikov:1996mc}. The choice of this cutoff is somehow arbitrary but it should exceed the typical momentum $k_*$ of the particles created at early times. For the simulations considered in this paper we will take $k_{\rm \Lambda}=4\, k_*$ with $k_*= 34.8\, M$ (see Appendix \ref{app2} for details). 

To clarify the role of perturbative decays in the delay of parametric resonance, we performed simulations both with and without the inclusion of fermions with model parameters $\xi = 1500$, $\lambda = 3.4 \times 10^{-3}$, $g_1^2=g_2^2=0.3$, $\alpha_{W^{\pm}}= 3\cdot 10^{-3}$ and $\alpha_Z= 5\cdot 10^{-3}$. The results are presented in Section \ref{sec2_1} and \ref{sec2_2}. 

\subsection{Boson production in the absence of  fermions} \label{sec2_1}

\begin{figure}
\subfigure[]{\includegraphics[scale=.39]{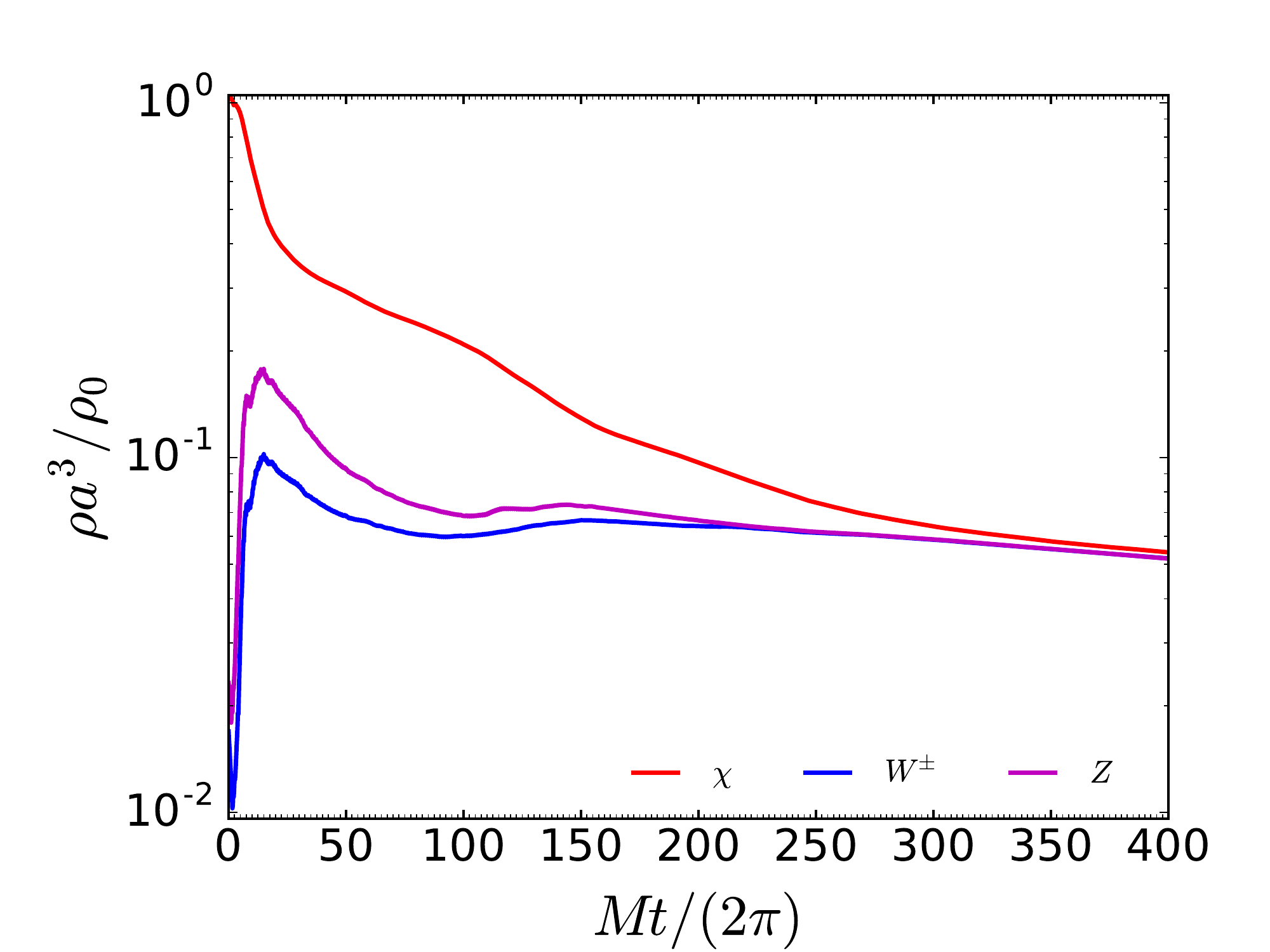}}
\subfigure[]{\includegraphics[scale=.39]{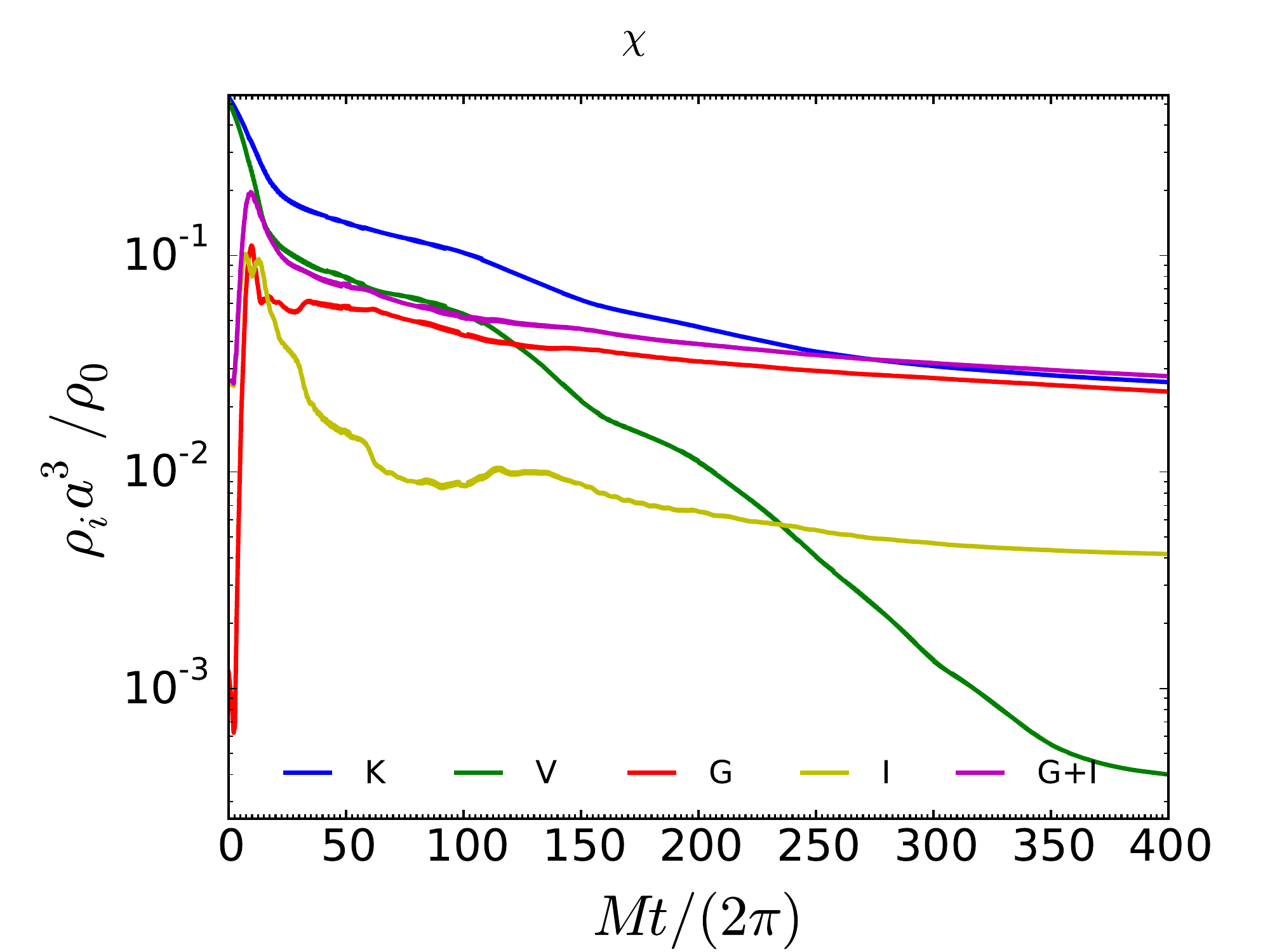}} 
\subfigure[]{\includegraphics[scale=.39]{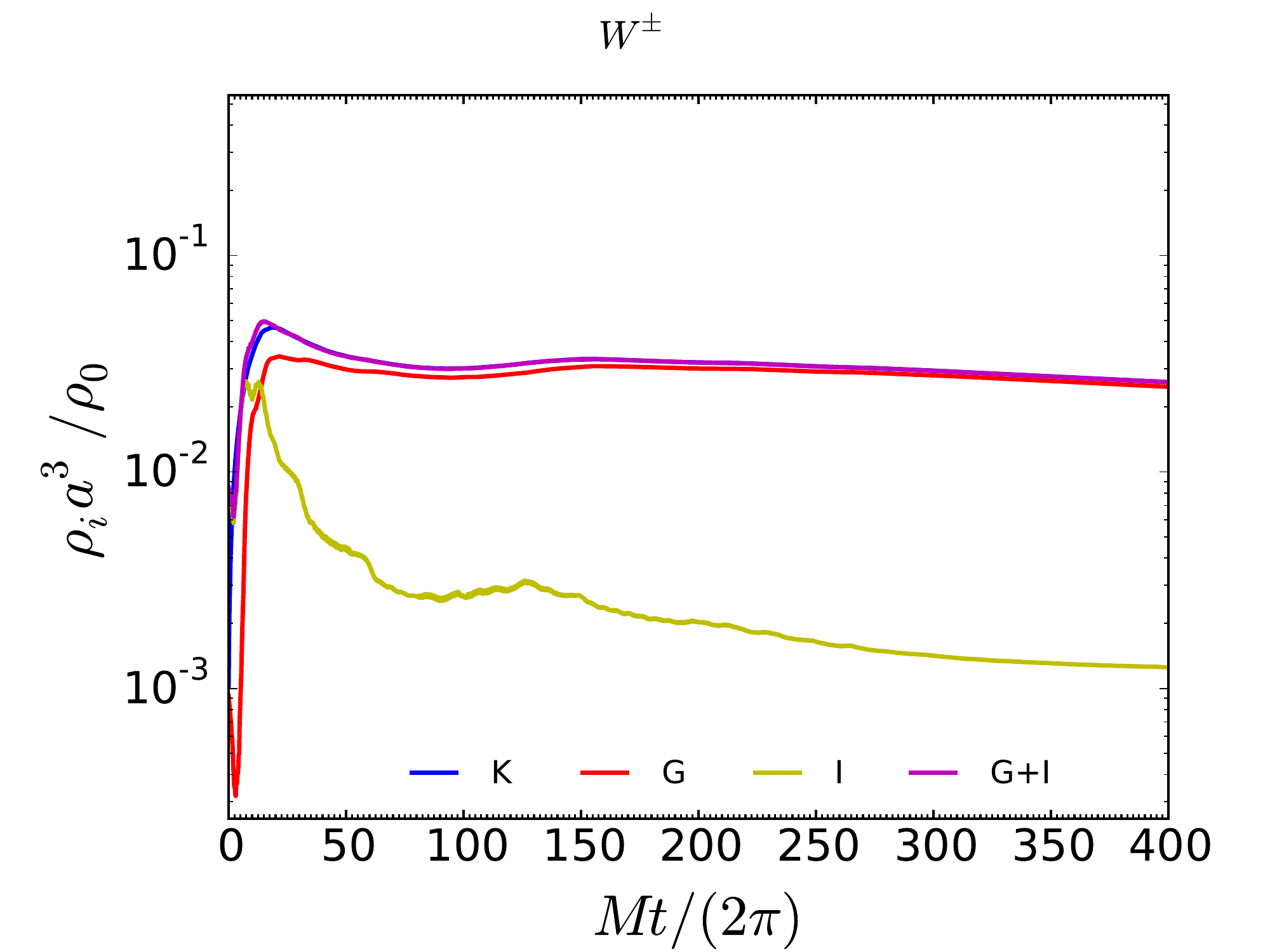}}
\subfigure[]{\includegraphics[scale=.39]{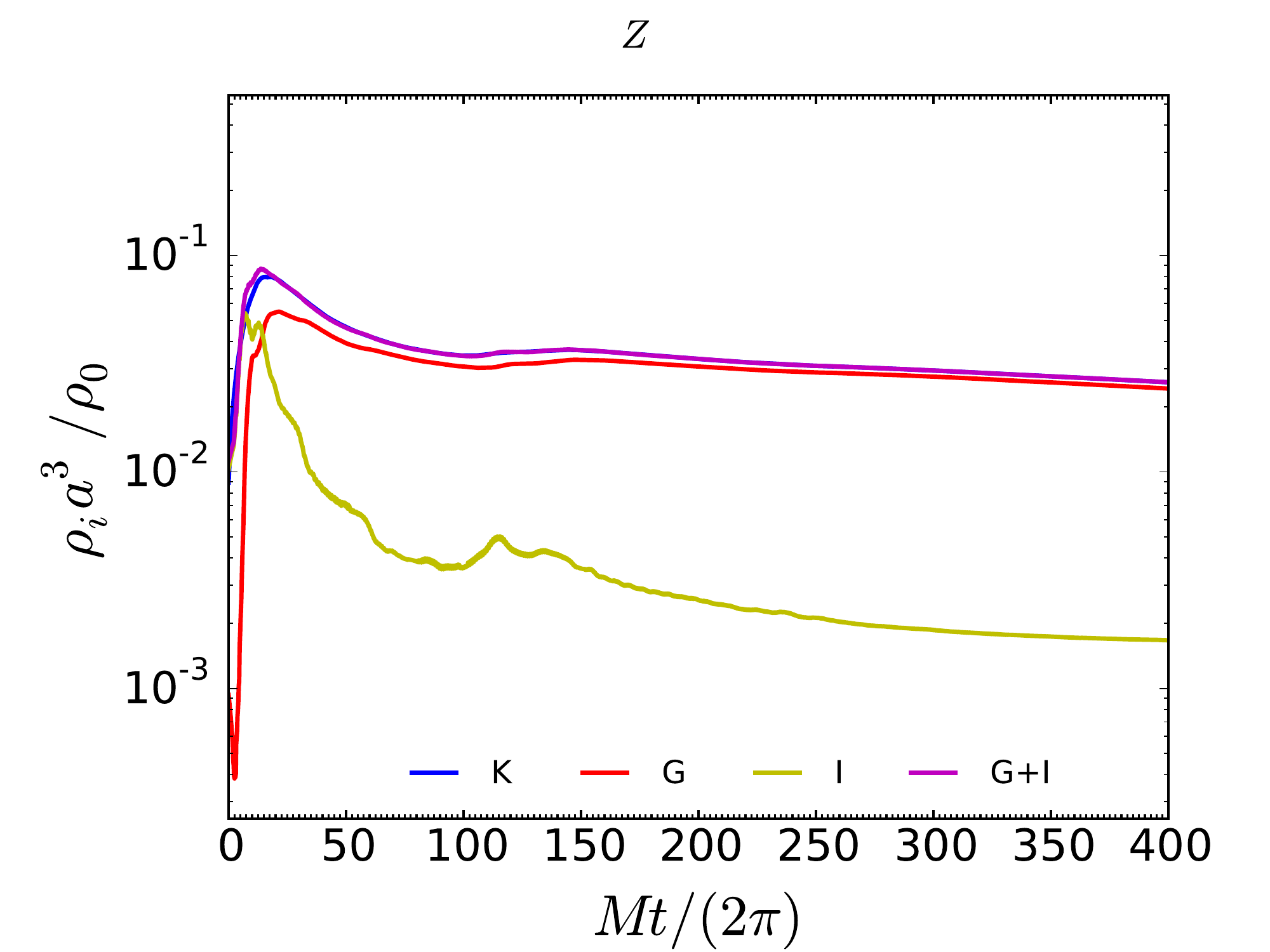}}
\caption{Evolution of the \textit{average} Higgs, $W^{\pm}$ and $Z$ energy densities in the absence of fermions [panel (a)] and the associated splitting into kinetic (K), potential (V), gradient (G) and interaction (I) contributions [panels (b) to (d)]. The normalization $\rho_0$ stands for the energy available at the onset of matter domination.}\label{totalE}
\end{figure}
\begin{figure}
\begin{center}
\subfigure{\includegraphics[scale=.375]{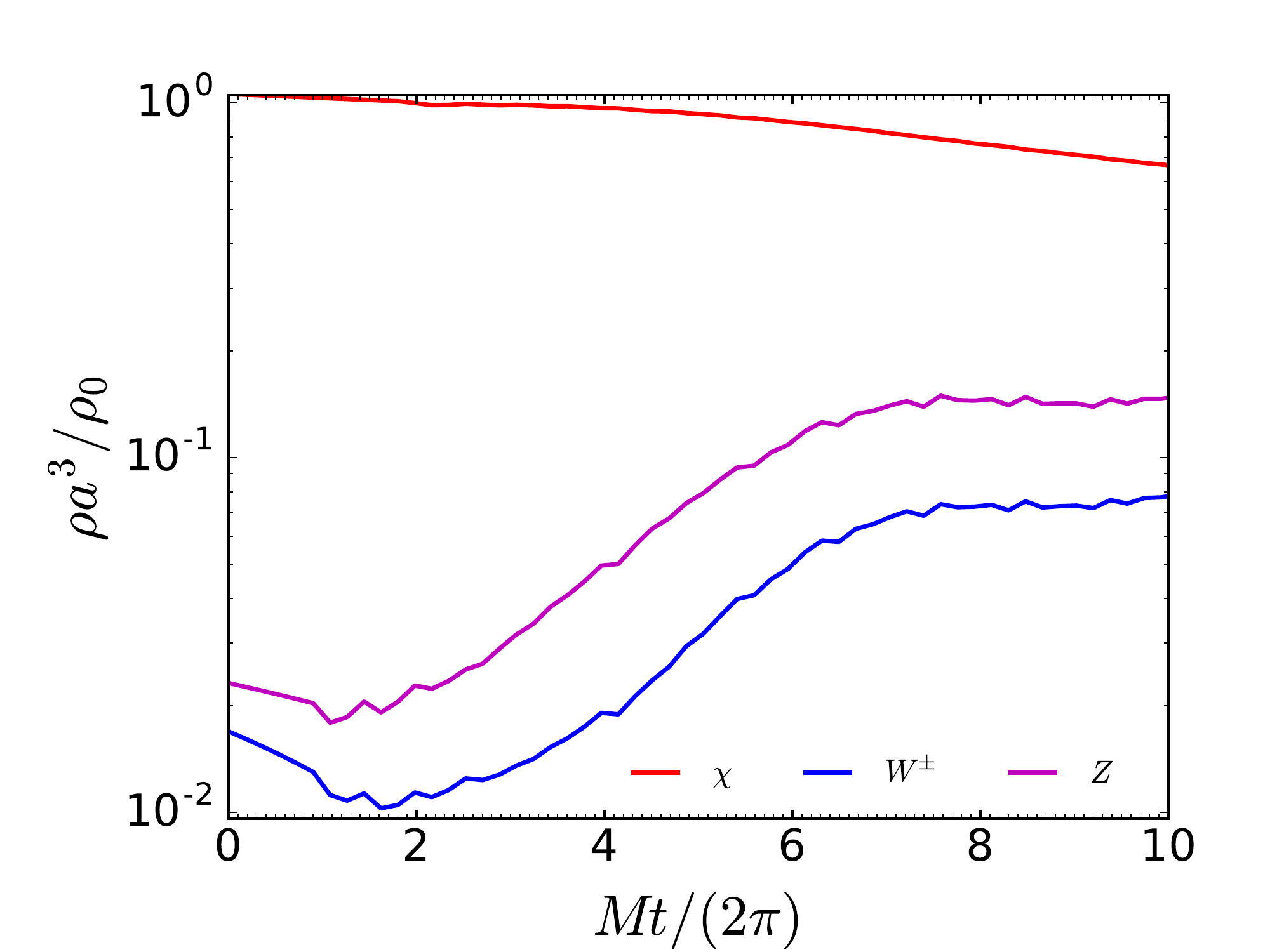}}
\subfigure{\includegraphics[scale=.375]{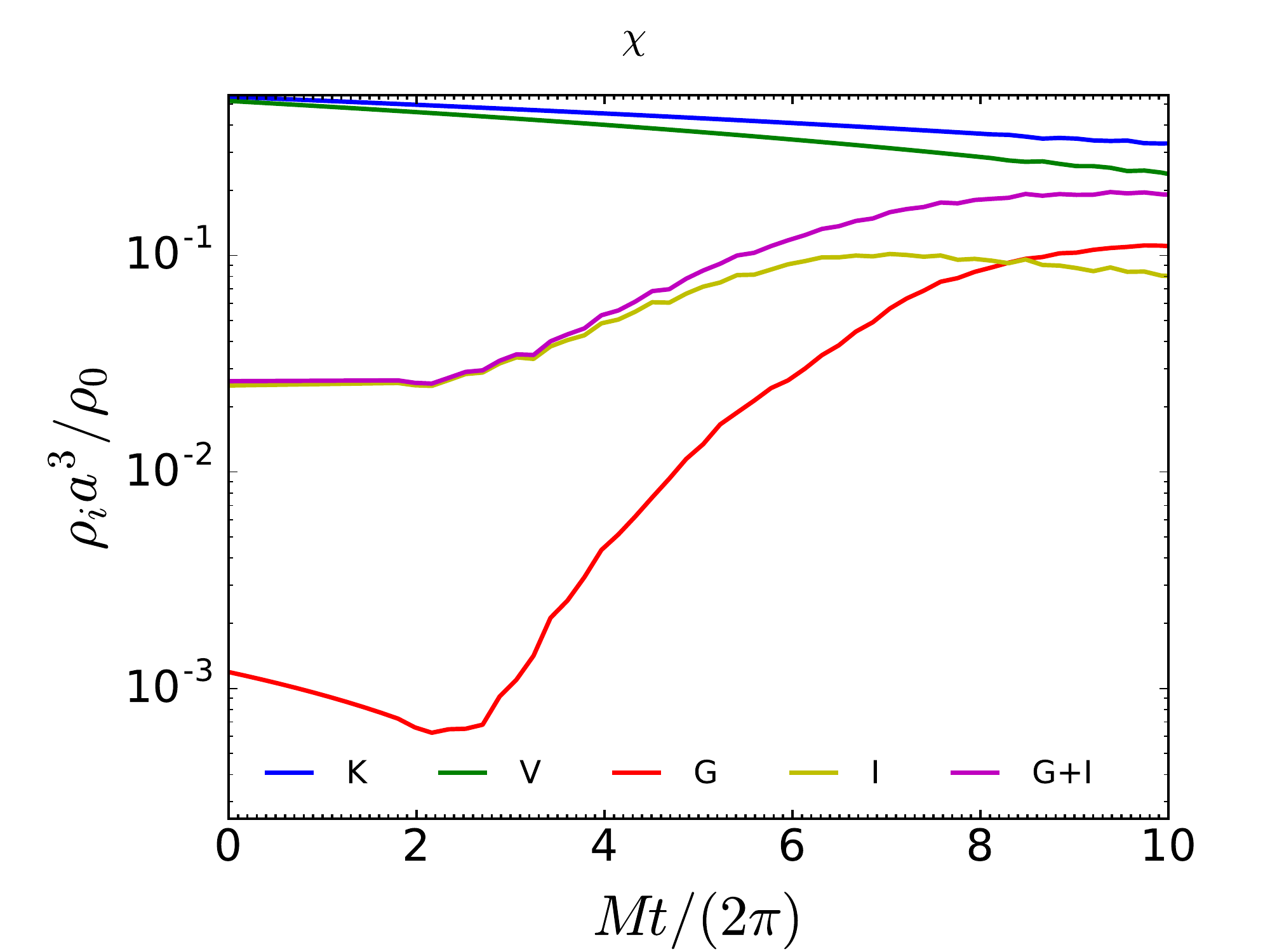}} 
\end{center}
\caption{(Left) Early time evolution of the \textit{average} Higgs, $W^{\pm}$ and $Z$ energy densities in the absence of fermions. (Right) Splitting of the Higgs energy density into kinetic (K), potential (V), gradient (G) and interaction (I) contributions in the same range. The normalization $\rho_0$ stands for the energy available at the onset of matter domination.}\label{totalEdetail}
\end{figure}

For solving the evolution equations \eqref{chieq}-\eqref{Feq} in the absence of fermions,\footnote{i.e. for $\Gamma_{B_i}=0$, $\rho_F=0$,  $\dot \rho_F=0$.} we use a lattice of length $L=0.50\, M^{-1}$ and $128^3$ points. This choice allows us to cover momenta in the range  $0.36\, k_*<k <40.23\, k_*$ during the whole simulation time $Mt/(2\pi)=400$. Once the field equations are solved, several observables such as the various components of the energy momentum tensor can be computed. In what follows, we focus on the evolution of the different energy densities and their spectral distributions:
\begin{enumerate}
\begin{figure}\vspace{-0.5cm}
\centering 
\subfigure{\includegraphics[scale=0.37]{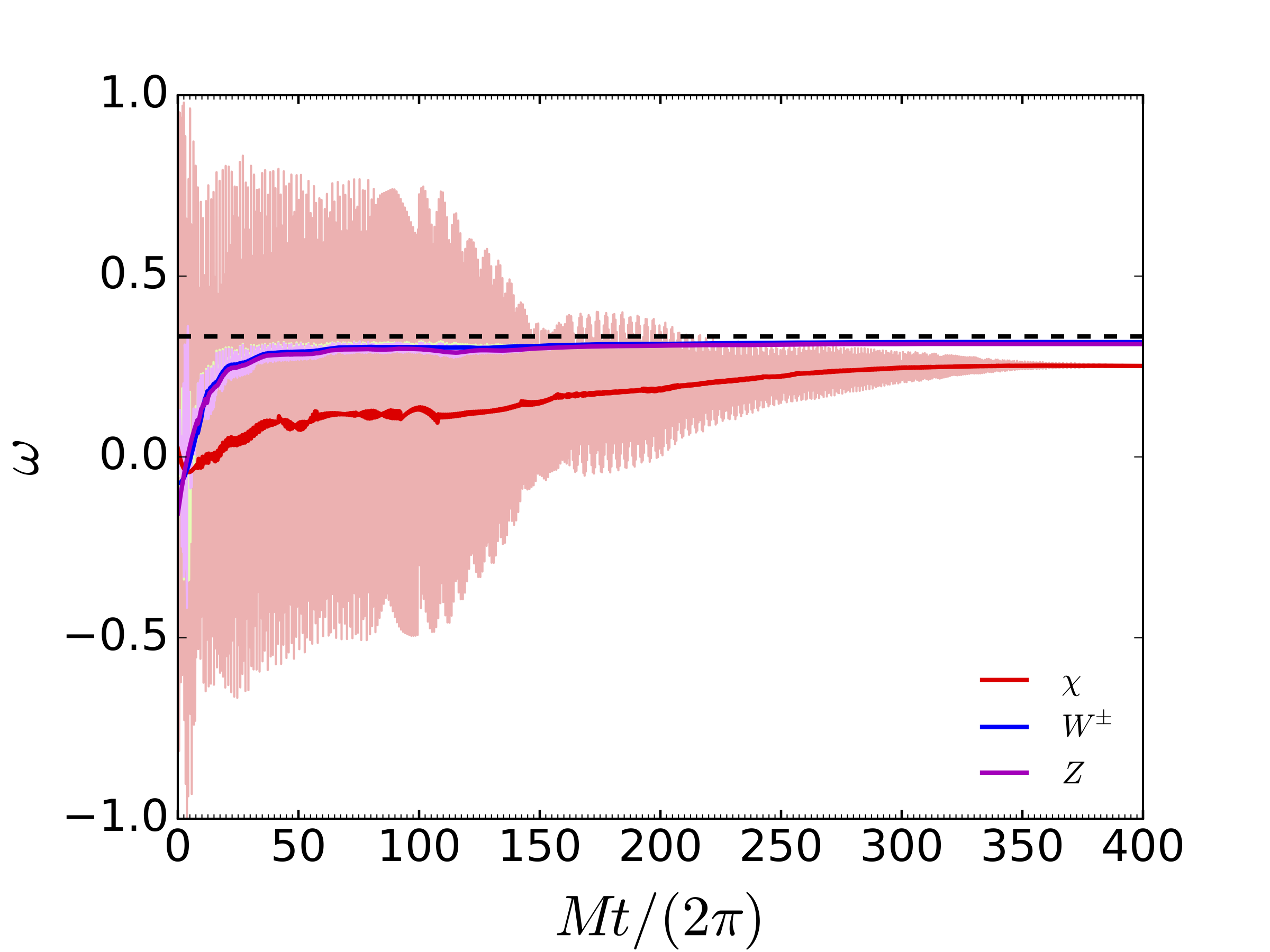}}
\subfigure{\includegraphics[scale=.37]{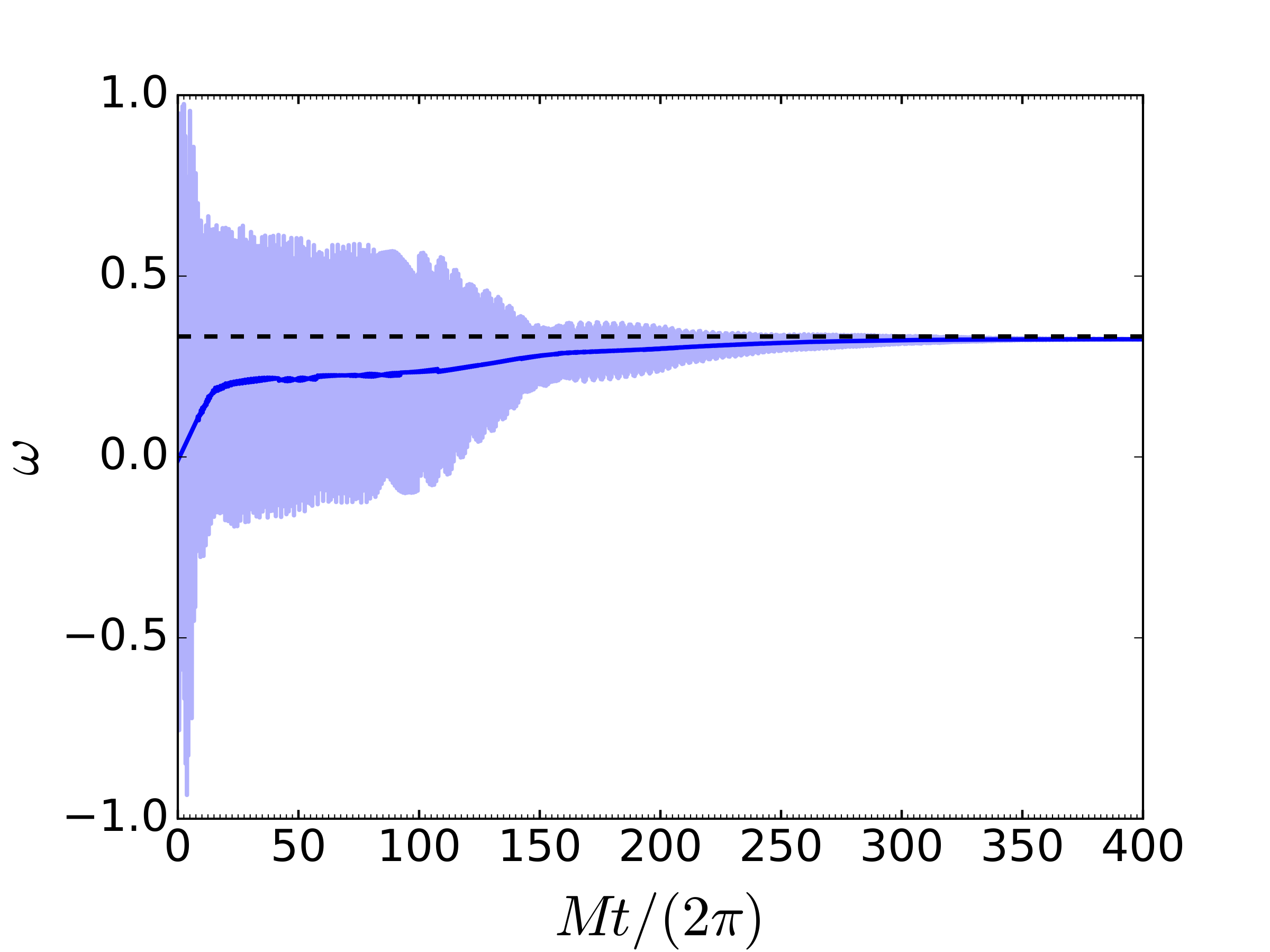}}
\caption{(Left) Evolution of the Higgs, $W^{\pm}$ and $Z$ equation-of-state parameters $w =\langle p \rangle/\langle \rho\rangle$ (shaded areas) and their mean values (solid lines) in the absence of fermions. (Right) Evolution of the global equation of state and its mean in the same case. The dashed black line corresponds to the ultrarelativistic limit $\omega = 1/3$.} \label{eqstate}
\end{figure}
\item{\bf Energy density:} The evolution of the \textit{average} Higgs, $W^{\pm}$ and $Z$ energy densities and their splitting into kinetic (K), potential (V), gradient (G) and interaction (I) contributions is presented in Fig.~\ref{totalE}. The sum of gradient and interaction (G+I) components is also included for later convenience. Fig.~\ref{totalEdetail} contains a zoom of the panels (a) and (b) in Fig.~\ref{totalE}. All quantities are normalized to the the total energy density available at the onset of matter domination and multiplied by the scale factor to the cube, whose initial value is taken to be $1$.

As shown in Fig.~\ref{totalEdetail}, the early dynamic of the system is driven by the oscillations of the zero mode around the minimum of the potential.  The kinetic and potential contributions of the Higgs field dominate over all other energy components and the Universe expands as non-relativisitic matter $(\rho_\chi a^{3}=\textrm{constant}$). 

 At each zero crossing of the Higgs field, a small fraction of energy is transferred into the $W^{\pm}$ and $Z$  bosons, which rapidly accumulate  due to bosonic stimulation effects and the absence of a depletion mechanism. This resonant production of 
 particles translates into a decay of the Higgs condensate  and an exponential growth of the gauge boson energy 
 densities.  As 
 shown in  panels (c) and (d) of Fig.~\ref{totalE}, all energy contributions are significantly excited during this stage. 
In agreement with the analytical expectations (cf. Appendix \ref{app2}), the production rate in 
 the absence of fermions \textit{does not depend} on the type of particle created at the bottom of the potential. $W^{\pm}$ and $Z$ bosons grow exactly at the same rate.
 
When the energy density of the created particles becomes comparable to the energy stored in the Higgs field ($ \rho_{B_i}/\rho_\chi \simeq  10\,\%  $, $Mt/(2\pi)\simeq 6$), the resonance terminates due to backreaction effects. From there on, the transfer of energy proceeds at a much smaller rate (see Fig.~\ref{totalE}). The remaining Higgs condensate acts as a localized source in the infrared which pumps energy into the system in a continuous and adiabatic way.  After some time, the potential energy of the condensate becomes smaller than the gauge boson energy density. When this happens, the kinetic and gradient contributions of the $B_i$ fields saturate to an almost constant amplitude (panels (c) and (d)), while the potential energy of the Higgs continues to decrease (panel (b)). Eventually, this component becomes smaller than the interaction contribution and the system approaches an equipartition distribution with $\textrm{K}\simeq \textrm{G}+ \textrm{I}$ (panels (b) to (d)). This asymptotic behavior can also be observed in the evolution of the different equation-of-state parameters (see Fig.~\ref{eqstate}), which change from $0$ to\footnote{Note that these values are close but not equal to $1/3$ due to the contribution of interactions to the equipartion equation $\textrm{K}\simeq \textrm{G}+ \textrm{I}$. The value $1/3$ would correspond to a case with K=G.}
\begin{equation}
w_{\chi} \simeq 0.25\,,\hspace{10mm} w_{W^{\pm}} \simeq 0.32\,,\hspace{10mm} w_{Z} \simeq 0.31\,,\hspace{10mm}  w_{T} \simeq 0.32\,,
\end{equation}
at $Mt/(2\pi)\simeq 400$. At that time, the total energy of the system $\rho_T$ is equally distributed among the different components
\begin{equation}\label{rhorationoF}
\frac{\rho_\chi}{\rho_T} \simeq 25.6\, \%\,,\hspace{10mm} \frac{\rho_{W^{+}}}{\rho_T} \simeq 24.8\, \%\,,\hspace{10mm} \frac{\rho_{W^{-}}}{\rho_T} \simeq 24.8\, \%\,,\hspace{10mm} \frac{\rho_Z}{\rho_T} \simeq 24.8\, \%\,.
\end{equation}
Note that the depletion of the Higgs field is not completely achieved in the absence of fermions. 
\begin{figure}
\centering 
\subfigure{\includegraphics[scale=0.37]{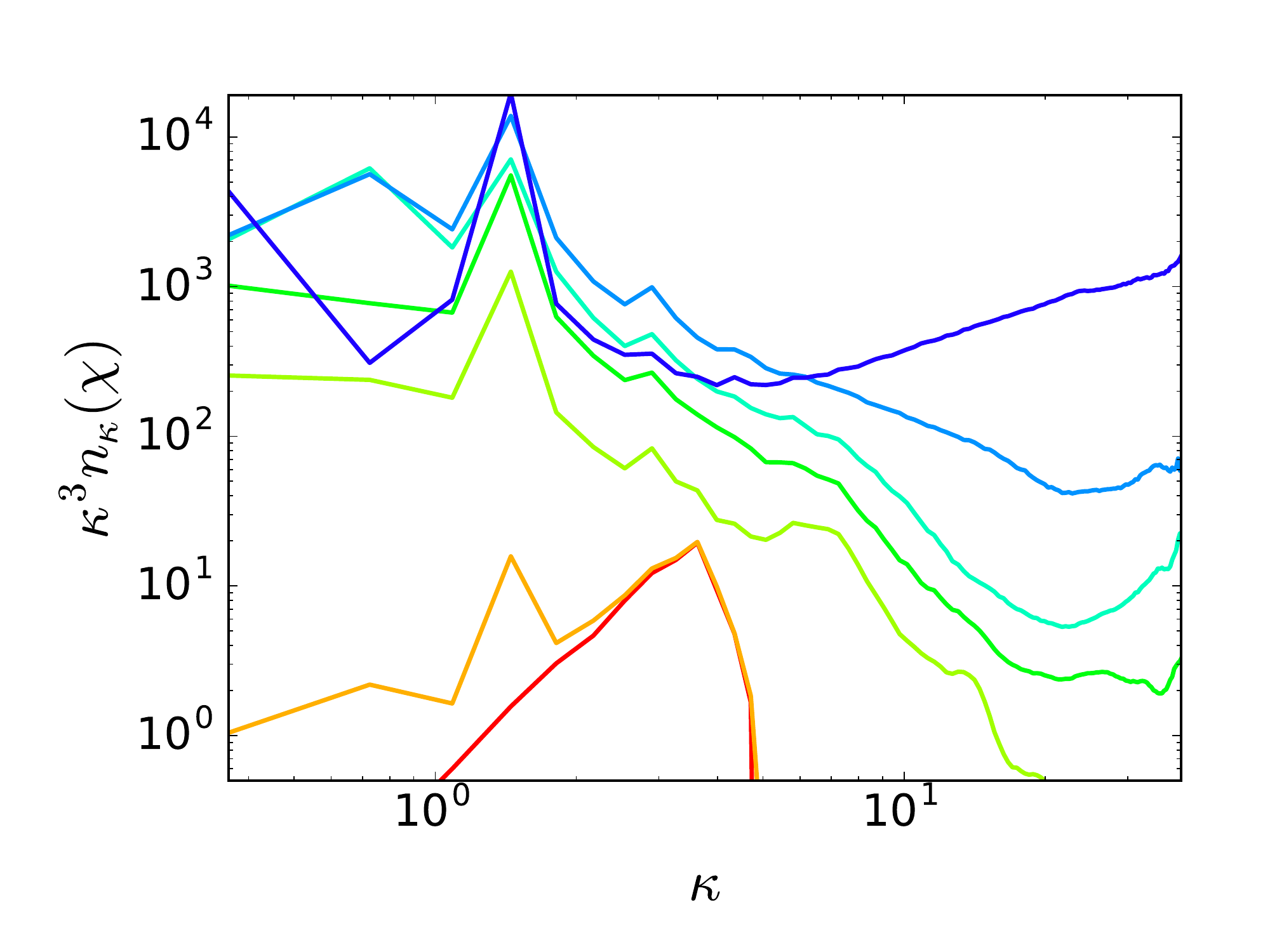}}
\subfigure{\includegraphics[scale=.37]{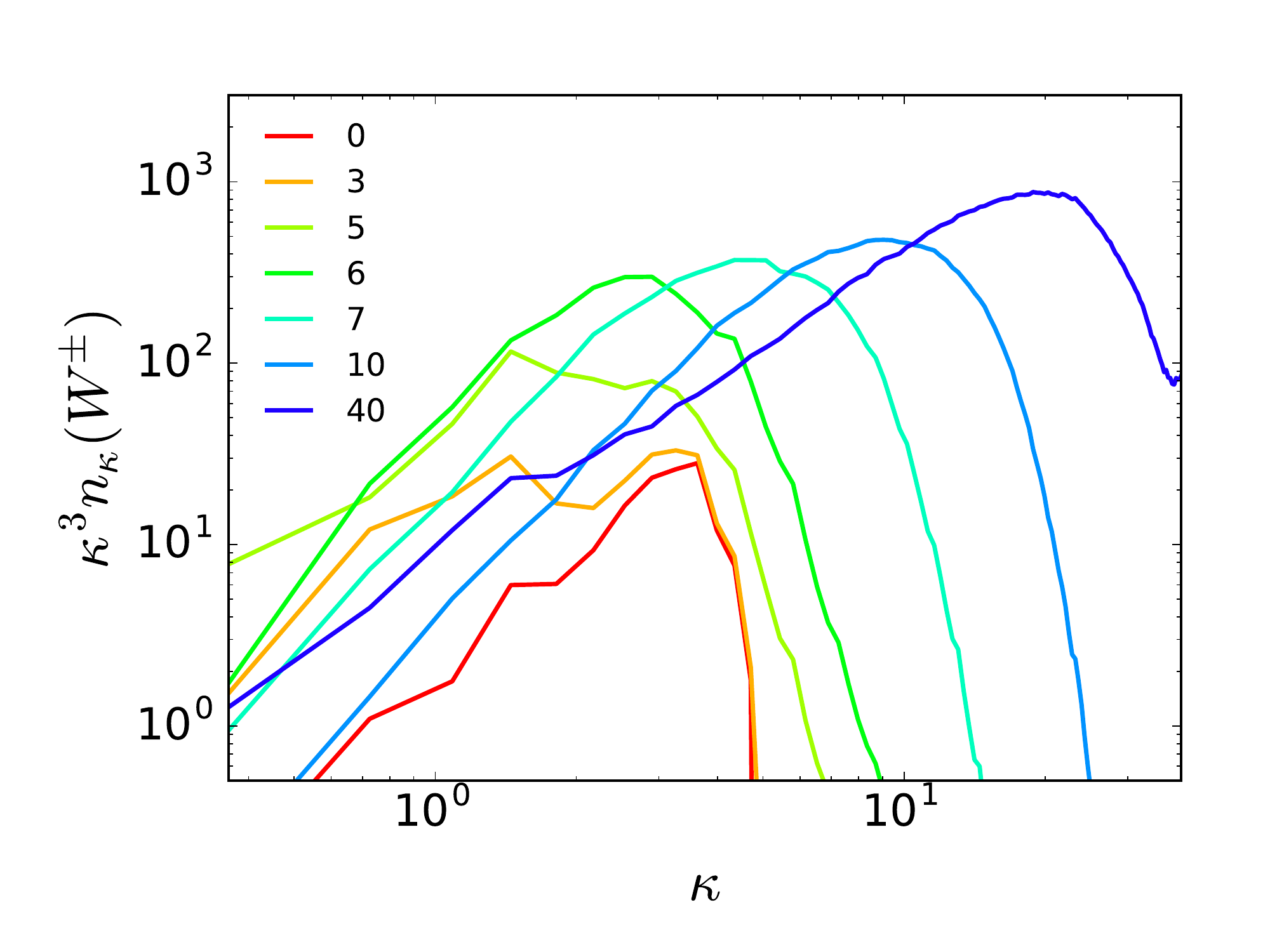}}
\subfigure{\includegraphics[scale=.37]{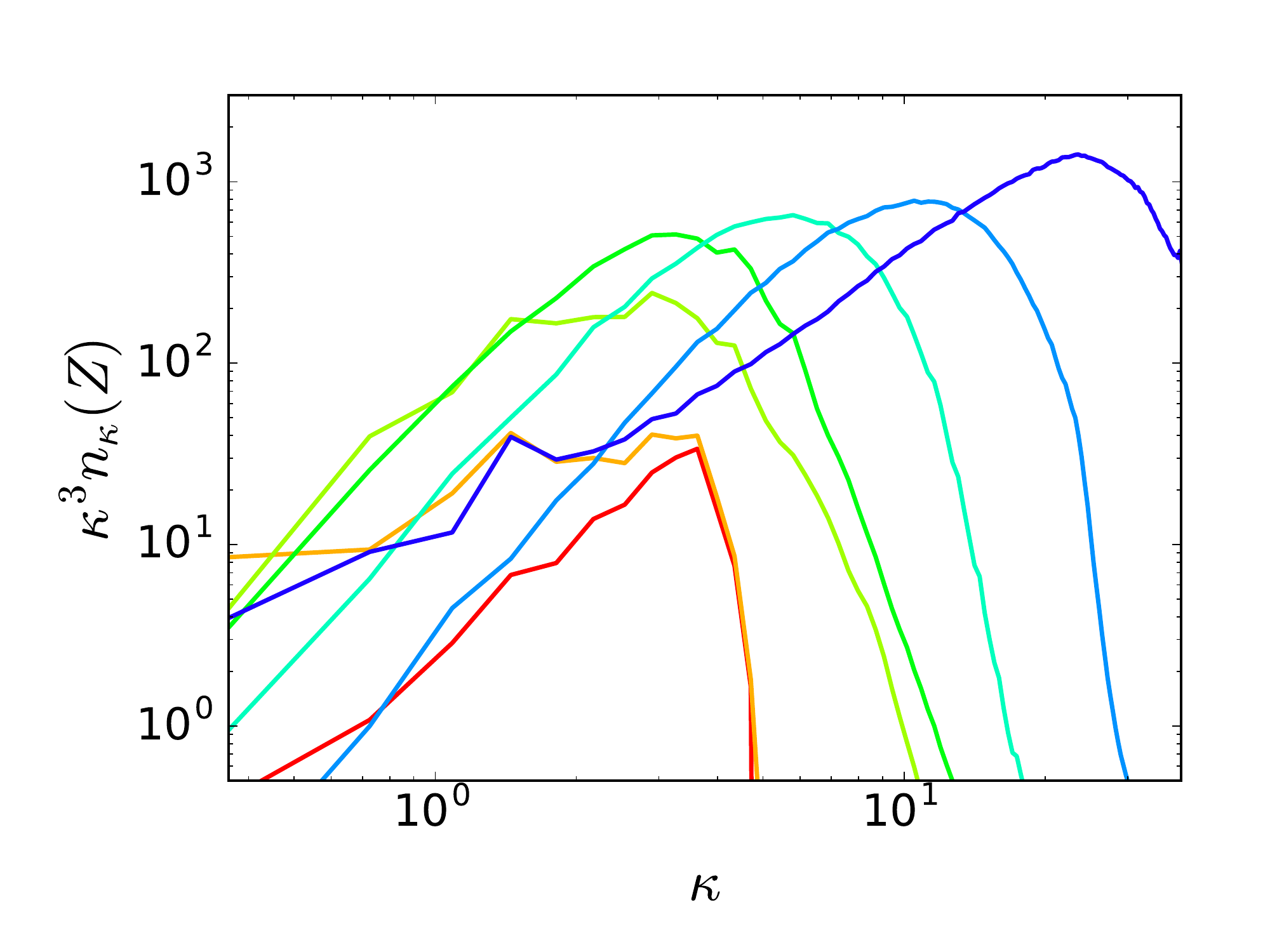}}
\caption{Spectral distributions $\kappa^3 n_\kappa$ for the Higgs, $W^{\pm}$ and $Z$ fields in the absence of fermions. 
Different colors correspond to different times (the precise values are indicated in the figures). 
The momenta $\kappa\equiv k/k_*$ are measured in units of the typical momentum $k_*$. Note that the actual 
power in the large $\kappa$ region of these figures is enhanced due to the $\kappa^3$ factor in 
$\kappa^3n_\kappa$. The occupation numbers for momenta close to the boundary of our simulation are not significantly 
populated during the whole simulation time. For details about the consistency checks regarding lattice artifacts 
cf. Sec.~\ref{sec2_3}.}\label{spectra}
\end{figure}
\item{\bf Spectra:} The evolution of the spectral distributions $\kappa^3 n_\kappa$  for the created Higgs, $W^{\pm}$ and $Z$ particles is shown in Fig.~\ref{spectra}. The momenta $\kappa\equiv k/k_*$ are measured in units of the 
typical momentum $k_*$. The diagonal lines at $t=0$ stand for the vacuum distributions below the cutoff scale $k_{\rm \Lambda}$. As expected, the early-time enhancements with respect to these vacuum distributions are restricted to momenta of order $\kappa \sim {\cal O}(1)$. The validity of the analytical estimates (see Appendix \ref{app2}) is limited to the very first oscillations, where the coupling between the different modes can be neglected. Indeed, rescattering effects become important soon after the beginning of the resonance and  give rise to a broadening of the spectra on top of the resonant amplification for $Mt/(2\pi)\gtrsim 5$.  As shown in Fig.~\ref{totalE}, the fraction of energy stored in the Higgs condensate at the end of parametric resonance is still rather large. The survival of this localized source of energy translates into the appearance of a 
turbulent regime  \cite{Micha:2002ey,Micha:2004bv} which will eventually lead to thermal equilibrium 
\cite{Berges:2013eia,Kurkela:2011ti}. The spectra in this regime are expected to cascade towards the
ultraviolet according to a self-similarity law.  Obtaining analytical estimates for this regime along 
the lines of  Refs.~\cite{Micha:2002ey,Micha:2004bv} is not an easy task however due to the non-polynomial
character of the interactions considered. 

The apparent power at large $\kappa$ in Fig.~4  is a consequence of the $\kappa^3$ scaling of 
$\kappa^3 n_\kappa$. When the modes in this region become populated, this scaling factor can be rather large 
($\sim10^4$), giving the impression that we are not properly accounting for lattice artifacts in our simulations. 
We performed however several consistency checks with respect to lattice artifacts that showed a weak dependence of 
the spectra on the precise choice of lattice parameters, cf. Sec.~\ref{sec2_3}.
\end{enumerate}

\subsection{Boson production in the presence of fermions} \label{sec2_2}
 
To solve Eqs.~\eqref{chieq}-\eqref{Feq} in the presence of  fermions we use a lattice cube of length $L=0.63\, M^{-1}$, $128^3$ points and minimum and maximum momentum coverage $k_{\rm min}=0.28\,k_*$ and $k_{\max}=31.61\,k_*$. As in Section \ref{sec2_1}, we focus on the behavior of the different energy densities and the associated spectral distributions:
\begin{figure}
\begin{center}
\includegraphics[scale=0.40]{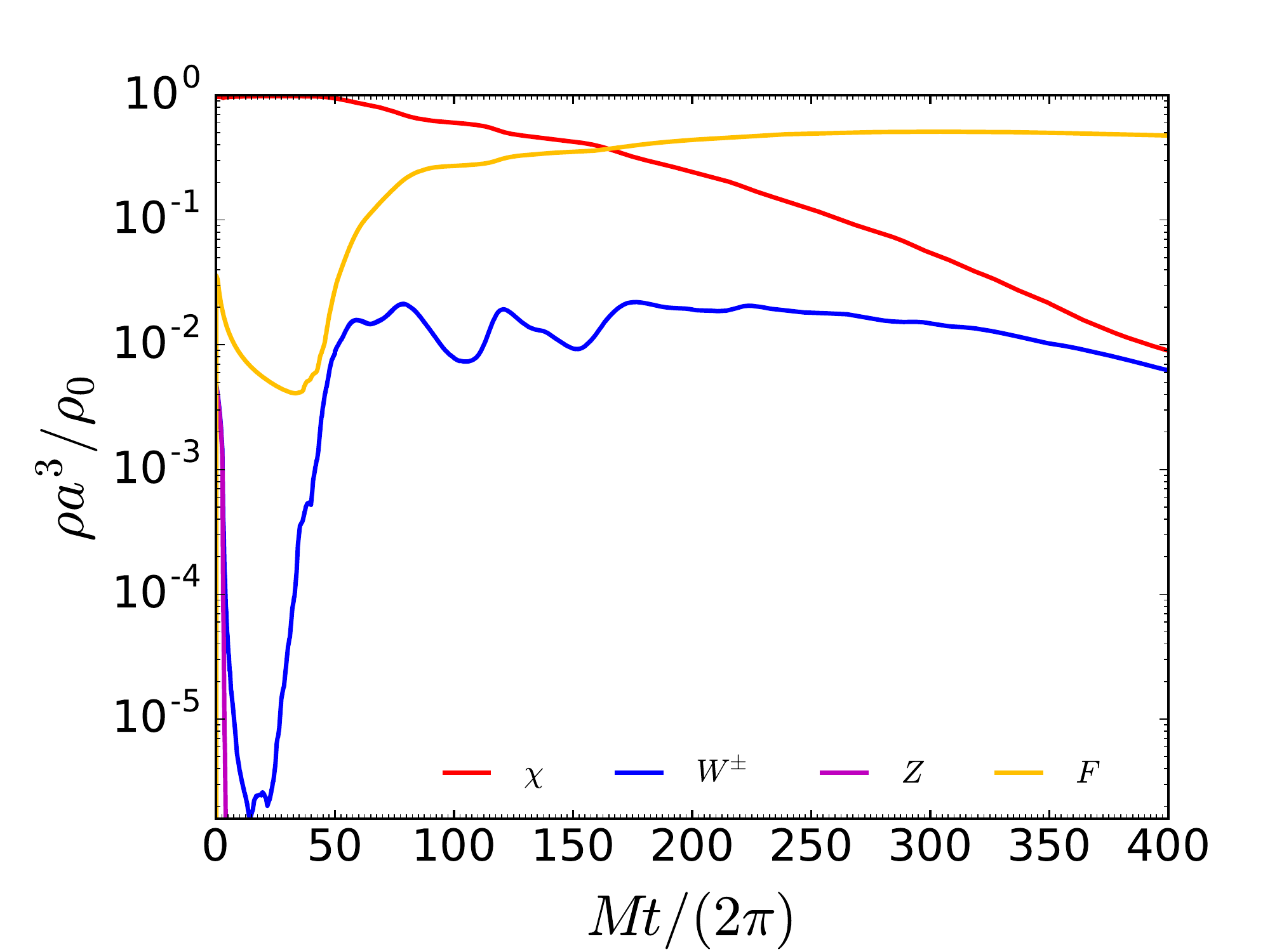}
\end{center}
\caption{Evolution of the \textit{average} Higgs, $W^{\pm}$ and $Z$ energy densities in the presence of fermions. The normalization $\rho_0$ stands for the energy available at the onset of matter domination.}\label{totalEF1} 
\end{figure}
\begin{enumerate}
\item {\bf Energy density:}  The evolution of the \textit{average} Higgs, $W^{\pm}$ and $Z$ energy densities strongly differs from that obtained in the absence of fermions. As shown in Fig.~\ref{totalEF1} and ~\ref{totalEF2}, the large expectation value of the Higgs field at the end of inflation induces a rapid decay of the gauge bosons created at the bottom of the potential and delays the onset of parametric resonance. In spite of the number of fermions created, their associated energy density stays completely subdominant with respect to the Higgs condensate and does not modify the matter-like expansion of the Universe  $(\rho_\chi a^{3}=\textrm{constant}$) for about  25 oscillations. Around this time, the situation changes dramatically.  While the $Z$ decay rate is still large, the decay rate of the $W^{\pm}$ bosons has  sufficiently decreased due to the expansion of the Universe as to allow these products to accumulate. Their 
energy density grows exponentially and eventually equals the energy density of fermions. When 
this happens ($Mt/(2\pi)\simeq 40$), the enhancement effect in the primary sector is transmitted to the fermion component, which  grows until $Mt/(2\pi)\simeq 60$. At this time, its energy density becomes comparable to that in the Higgs field ($\rho_F/\rho_\chi \simeq  10\,\%$)  and the fast transfer of energy terminates due to backreaction effects.  The dragging of energy out of the condensate continues through a stage of turbulence. As shown in Fig.~\ref{eqstateF}, this turbulent regime gives rise to a slow evolution of the effective $W^{\pm}$ equation of state towards a value significantly different from zero 
\begin{equation}
w_{\chi} \simeq 0.22\,,\hspace{10mm} w_{W^{\pm}} \simeq 0.31\,,\hspace{10mm}  w_{T} \simeq 0.33\,,
\end{equation}
at $Mt/(2\pi)\simeq 400$. At that time, the total energy density of the Universe $\rho_T$ is completely dominated by the fermionic degrees of freedom
\begin{equation}
\frac{\rho_\chi}{\rho_T} \simeq1.8\, \%\,,\hspace{10mm} \frac{\rho_{W^{+}}}{\rho_T} \simeq 1.2\, \%,\hspace{10mm} \frac{\rho_{W^{-}}}{\rho_T} \simeq 1.2\, \%,\hspace{10mm} \frac{\rho_F}{\rho_T} \simeq 95.8\, \%\,.
\end{equation}
The comparison of the above ratios with those obtained in the absence of decays (cf. Eq.~\eqref{rhorationoF}) makes apparent the central role of fermions in the depletion of the inflaton field.
\begin{figure}
\subfigure{\includegraphics[scale=.38]{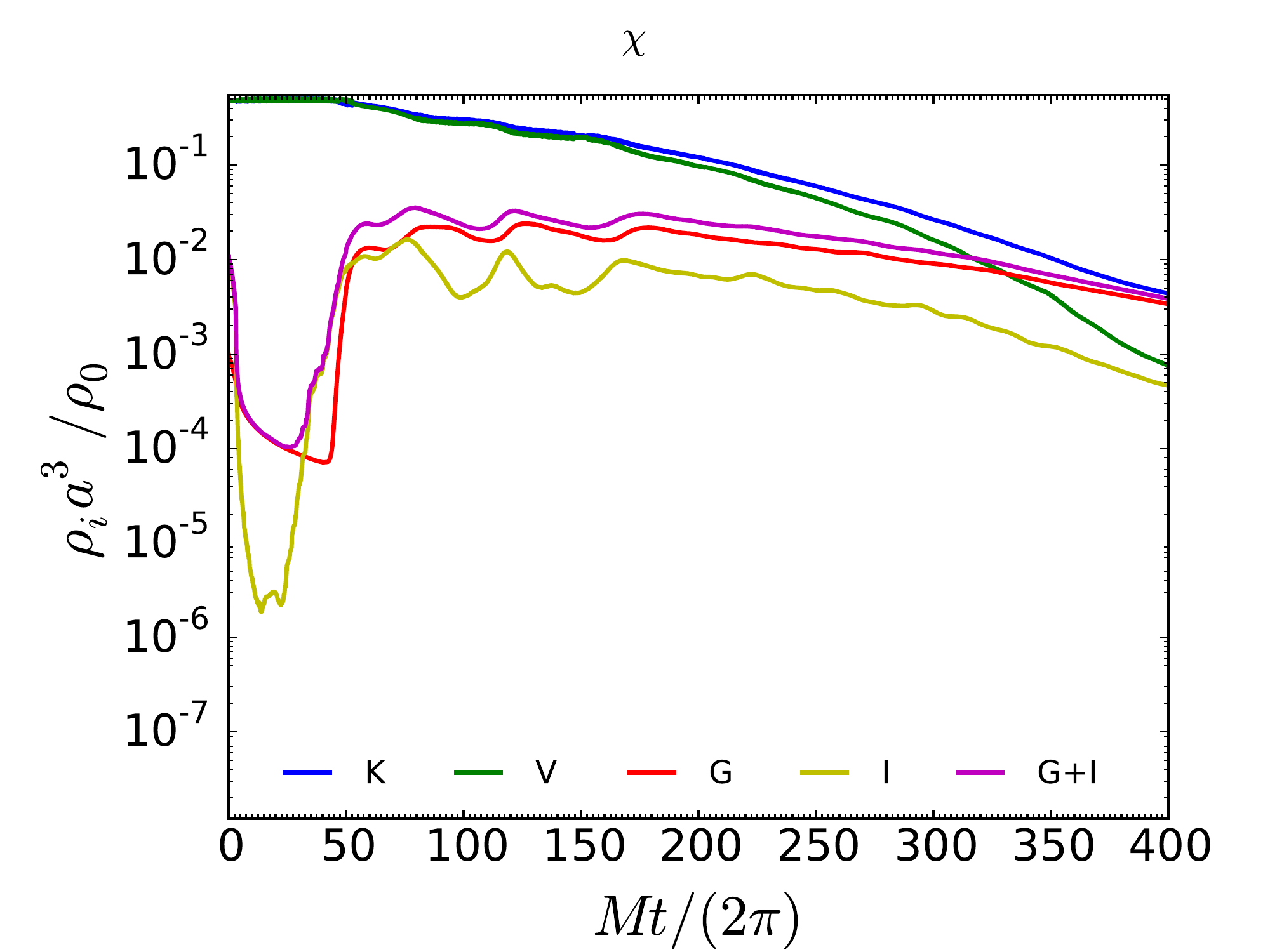}} 
\subfigure{\includegraphics[scale=.38]{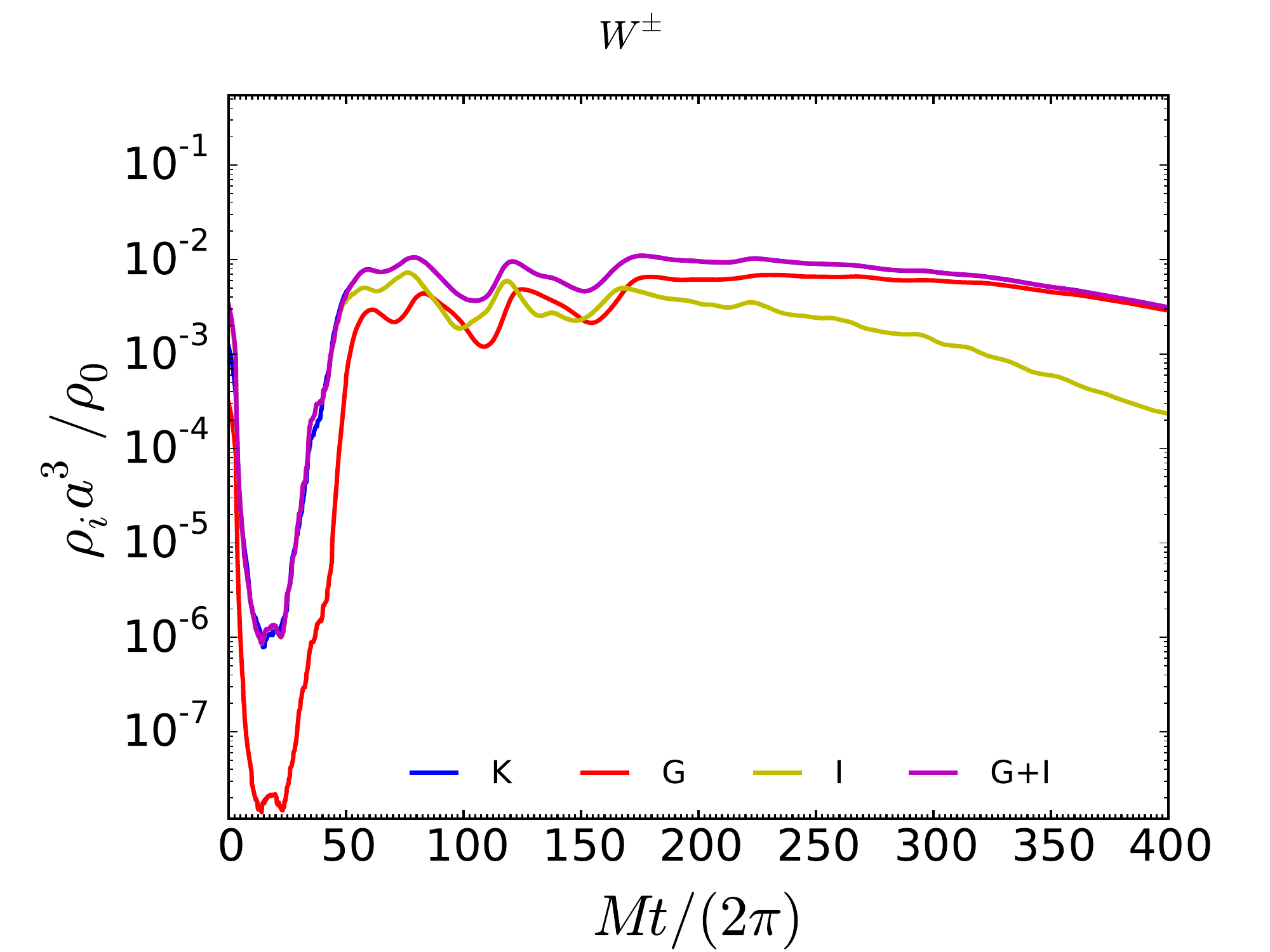}}
\caption{Splitting of the Higgs and $W^{\pm}$ energy densities into kinetic (K), potential (V), gradient (G) and interaction (I) contributions in the presence of fermions. The normalization $\rho_0$ stands for the energy available at the onset of matter domination.}\label{totalEF2}
\end{figure}
\begin{figure}
\centering 
\subfigure{\includegraphics[scale=0.37]{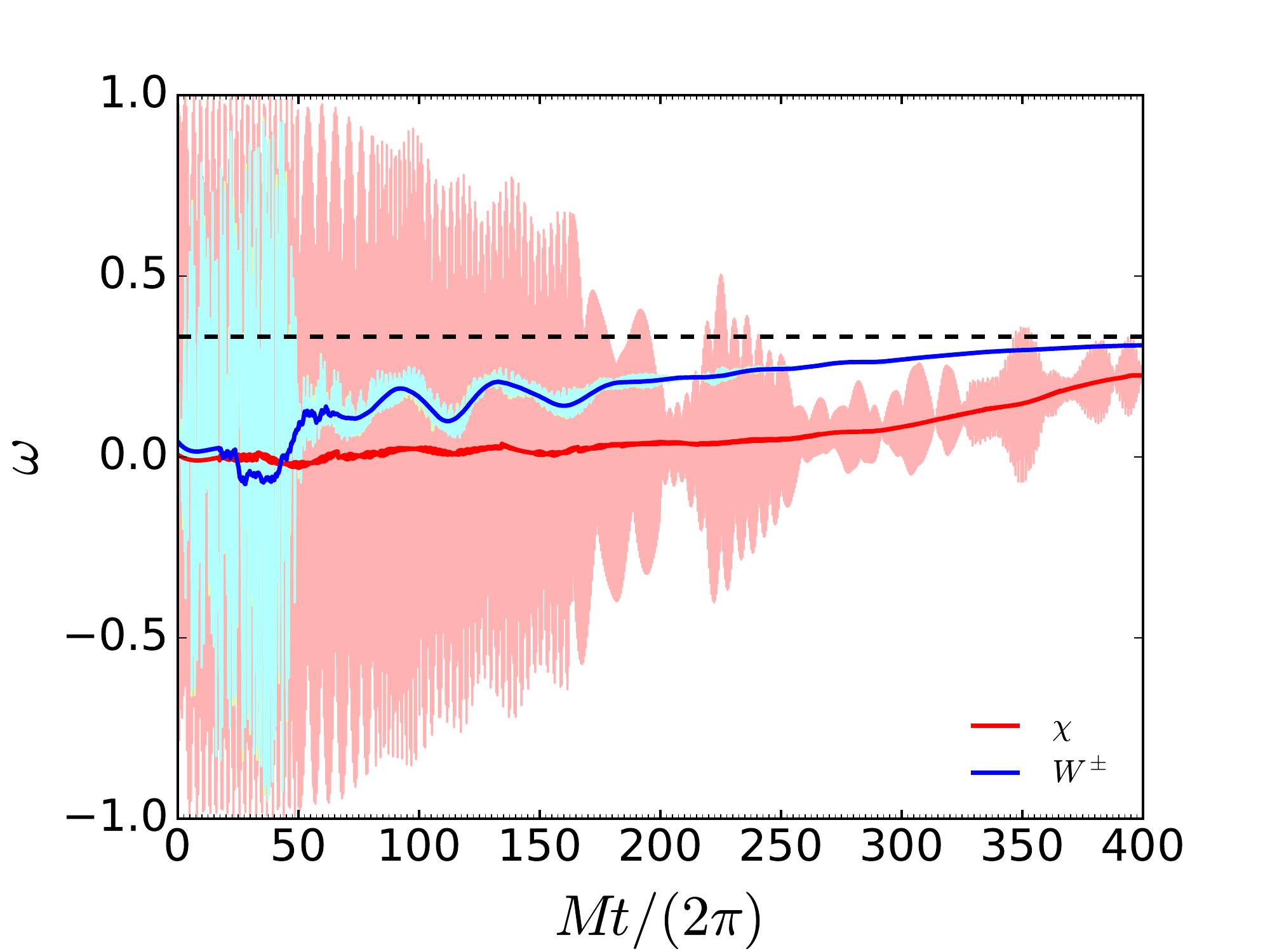}}
\subfigure{\includegraphics[scale=.37]{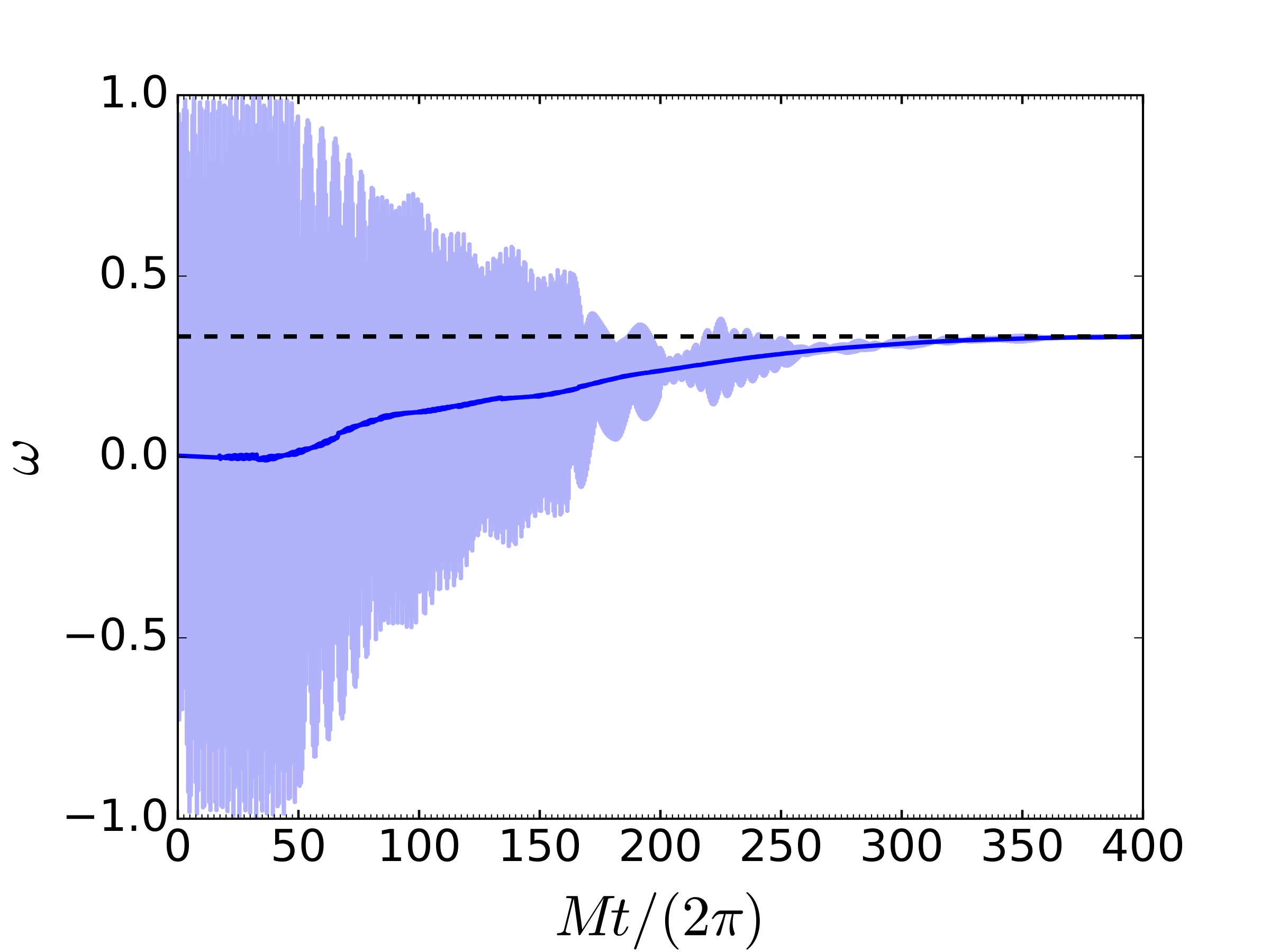}}
\caption{(Left) Evolution of the Higgs, $W^{\pm}$ and $Z$ effective equation-of-state parameters $w =\langle p \rangle/\langle \rho\rangle$ (shaded areas) and their mean values (solid lines) in the presence of fermions. (Right) Evolution of the global equation of state and its mean in the same case. The dashed black line corresponds to the ultrarelativistic limit $\omega = 1/3$.}\label{eqstateF}
\end{figure}
\item {\bf Spectra:} The evolution of the spectral distributions for the created Higgs and $W^{\pm}$ bosons is shown in Fig.~\ref{spectraF}.  Since $Z$ particles are not significantly produced during the whole simulation, we  decided to omit the corresponding spectrum. In agreement with the evolution of the energy densities presented above, the occupation numbers of the Higgs and $W^{\pm}$ bosons remain small during the first tens of oscillations.  The distributions stay below $\kappa \sim {\cal O}(1)$ during the initial stages of the resonance ($25 \lesssim Mt/(2\pi)\lesssim40$). From there on, they evolve towards higher momenta due to rescattering effects. 
\end{enumerate} 
\begin{figure}
\centering 
\subfigure{\includegraphics[scale=0.37]{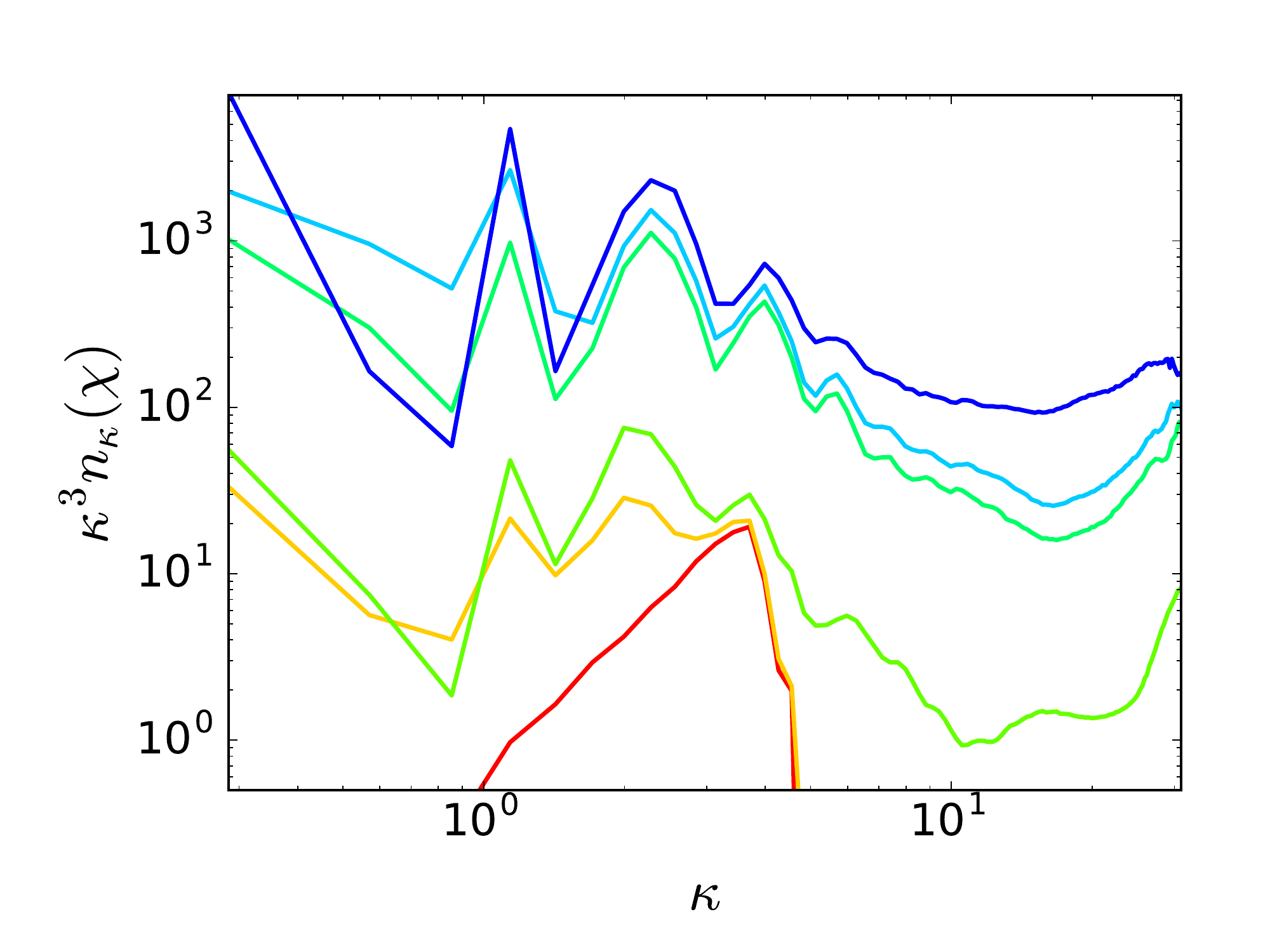}}
\subfigure{\includegraphics[scale=.37]{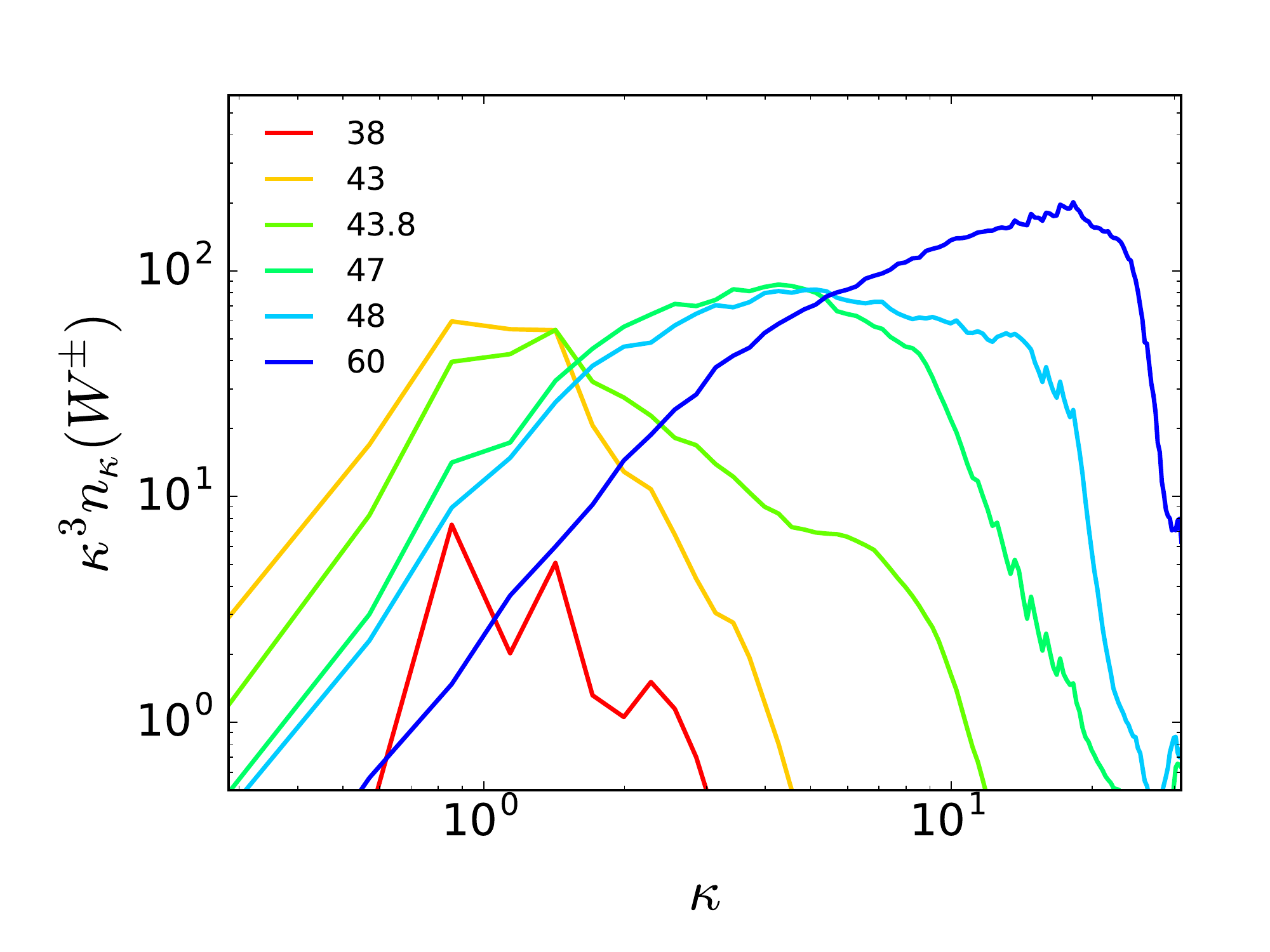}}
\caption{Spectral distributions $\kappa^3 n_\kappa$ for the Higgs and the $W^{\pm}$ fields in the presence of fermions. Different colors correspond to different times (the precise values are indicated in the figure). The 
momenta $\kappa\equiv k/k_*$ are measured in units of the typical momentum $k_*$. 
Note again that the actual power in the large $\kappa$ region of these figures is enhanced due to the $\kappa^3$ factor in 
$\kappa^3n_\kappa$. The occupation numbers for momenta close to the boundary of our simulation are not significantly 
populated during the whole simulation time. Consistency checks with respect to lattice 
artifacts are presented in Sec.~\ref{sec2_3}.}\label{spectraF}
\end{figure}

\section{Robustness of the numerical results}\label{sec2_3}
 
We used the redundancy of the equations \eqref{chieq}-\eqref{FRW} as a check of our numerical simulations. The 
results presented in this paper are consistent with energy conservation at the ${\cal O}(10^{-3}-10^{-2})$ level.

\begin{figure}
\subfigure{\includegraphics[scale=.39]{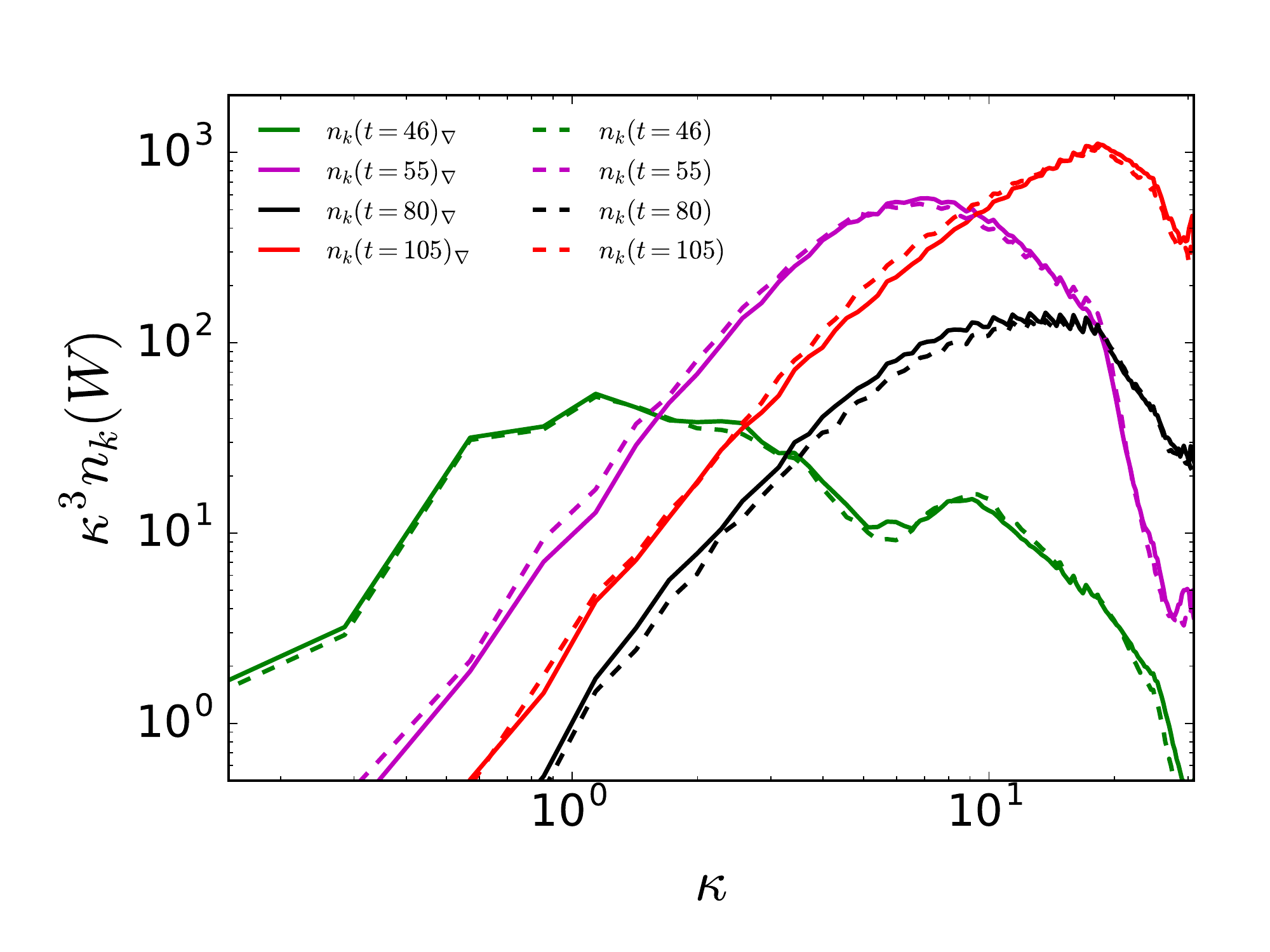}}
\subfigure{\includegraphics[scale=.39]{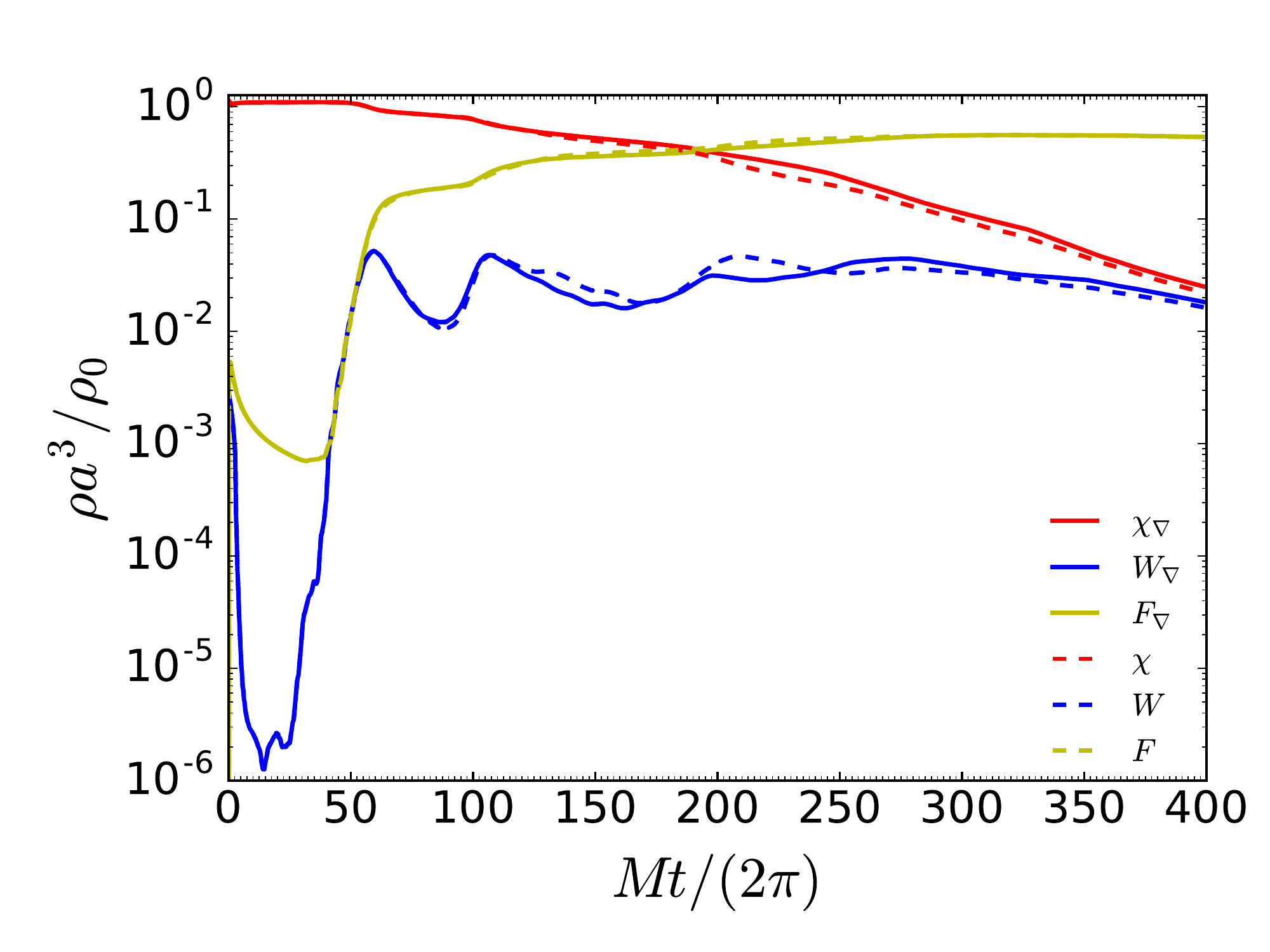}}
\caption{(Left) The spectra of the $W$ boson at different times computed with and without the gradient terms in the 
right-hand side of Eq.~\eqref{Feq}. The subindex $\nabla$ corresponds to the case with gradients. (Right) Evolution 
of the energy density of the different species in the same cases. 
\label{fig:GRADComparison}}
\end{figure}

We intentionally neglected the boundary terms on the right-hand side of Eq.~\eqref{Feq}. This decreases the computational 
time and avoids numerical errors coming from the finite differences approximation of the gradient. We performed tests 
including and excluding these terms to verify the numerical validity of this analytically consistent approximation. Given 
the computational time required by these consistency checks, we decided to include only one of the three gauge 
bosons $B_i$, namely one of the $W$ fields.\footnote{This is enough for our purposes. The two $W$ bosons evolve in a similar 
way and the $Z$ boson is not significantly produced in the presence of fermions.} The left panel of Fig.~\ref{fig:GRADComparison}
compares the spectral distributions of this component at different times in the presence and absence of boundary terms. No
significant modification on the spectra during the whole simulation time is observed. The evolution of the energy densities 
shown in the right panel confirms this fact.

\begin{figure}
\begin{center}
\includegraphics[scale=.39]{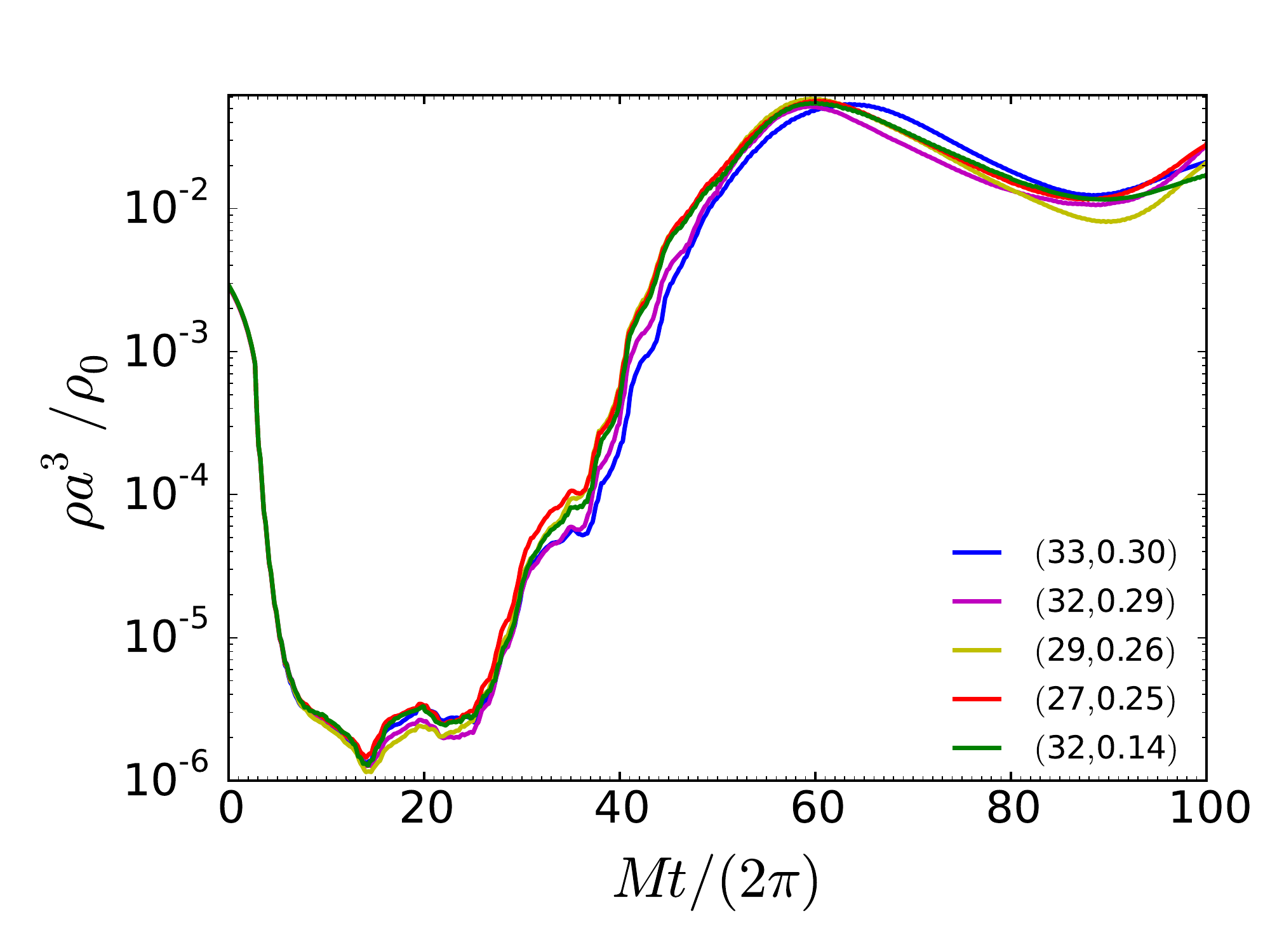}
\caption{Evolution of the total energy density of 
the $W$ bosons for different sets of parameters ($\kappa_{\text{max}},\kappa_{\text{min}}$). 
The magenta line corresponds to the particular 
choice of parameters used in the paper.\label{fig:ComparisonFermions}}
\end{center}
\end{figure}

The lattice parameters used in Sections \ref{sec2_1} and \ref{sec2_2} were chosen to ensure a sufficient coverage of the small 
momenta at which the particles are produced and the high momenta that become populated when the system becomes highly non-linear.
To verify the robustness of our results versus lattice artifacts we performed simulations with different choices of parameters.  
As above, we decided to include only one $W$ boson to save computational time. In spite of the better or worse coverage in momenta, 
the main physical aspects of the \textit{Combined Preheating} scenario were sufficiently captured in all the simulations. 
In particular, the evolution of the $W$ boson energy density presented in Fig.~\ref{fig:ComparisonFermions} shows 
little dependence on ($\kappa_{\text{max}},\kappa_{\text{min}}$).


\section{Scope and limitations of the results}\label{sec3}

The numerical results presented in Sections \ref{sec2_1} and \ref{sec2_2} confirm the qualitative expectations of the \textit{Combined Preheating} scenario and shed some light on the limitations of the analytical techniques. The formalism used in Refs.~\cite{GarciaBellido:2008ab,Bezrukov:2008ut} seems to be an appropriate way of estimating the number of oscillations needed to enter the parametric resonance regime but fails to determine the time  at which the energy of the secondary species exceeds the energy of the inflaton. As shown in Fig.~\ref{totalEF1}, when the energy density into the secondary species becomes comparable to that in the Higgs field ($\rho_F/\rho_\chi \simeq  10\,\%$), the resonance terminates and the transfer of energy continues at a much smaller rate, characteristic of a turbulent regime. The temporal scale of reheating is not determined by parametric resonance but rather by the approach to equipartition.

The choice of Higgs inflation in this paper should be understood as the choice of a working arena for understanding the \textit{Combined Preheating} mechanism in the non-linear regime. We performed several approximations to reduce the physical scenario to a baseline model that could be treated in a simple way with analytical and numerical techniques. Due to these approximations, the extrapolation of our results to the realistic Higgs inflation scenario should be done with care. In particular, one should not try to extract quantitative information about physically relevant parameters such as the depletion time of the condensate or the reheating temperature without properly taking into account the following issues: 

\begin{itemize}

\item \textit{Gauge boson polarizations}: We did not explicitly consider the three polarizations of the intermediate SM gauge bosons. However, these could be easily taken into account by replacing the $q_{B_i}$ and $\alpha_{B_i}$ parameters by effective parameters encoding this fact.  

\item  \textit{Gauge structure of the Standard Model}: We replaced the gauge interactions among the Higgs field and the 
SM gauge bosons by global interactions with three scalar fields $B_i$. The production of particles in models involving
gauge interactions have been widely studied in the literature (see, for instance,
\cite{Deskins:2013lfx,  GarciaBellido:2002aj,DiazGil:2007dy,DiazGil:2008tf,Figueroa:2015rqa,Rajantie:2000fd,
Enqvist:2014tta,Enqvist:2015sua,Lozanov:2016pac}). As explicitly demonstrated in Ref.~\cite{Figueroa:2015rqa}, the qualitative behavior of models involving global and purely Abelian interactions is very similar. The situation becomes more complicated when one considers non-Abelian interactions. 
These interactions modify the effective $B_i$ masses by adding extra non-linearities on top of the Higgs-gauge boson couplings.  Although these non-linearities are certainly small during the first stages of parametric resonance, they might become important when the number of created particles becomes large.

The global analogues of non-Abelian $SU(2)$ self-interactions $B_i^3,  B_i\partial B_i$ and $\partial B_i^2$ were completely ignored in our simulations. The impact of the omitted pieces can be estimated by computing, within our simulations, the ratio between the effective masses generated by the gauge self-interactions  and the Higgs-given masses \eqref{masses}. In the Hartree approximation, this ratio is of order\footnote{Note that there is no quartic $Z$ boson coupling in the Standard Model.}
\be\label{nonAratio}
\Delta_{B_i}\equiv  \frac{g_2^2\langle W^2\rangle }{\tilde m_{B_i}^2}\,.
 \ee 
As shown in Fig.~\ref{ratioDelta}, the impact of trilinear interactions on the oscillation frequency 
of the $B_i$ fields is expected to be moderate. The maximum value of $\Delta_{B_i}$ is achieved when 
the energy density into gauge bosons becomes comparable to the energy density of the Higgs or 
fermion components ($ \rho_B,\rho_F\simeq  0.1\rho_\chi$). At that time particle production stops
due to backreaction effects and $\Delta_{B_i}$ decreases significantly. 

\begin{figure}\vspace{-0.5cm}
\centering 
\subfigure{\includegraphics[scale=0.37]{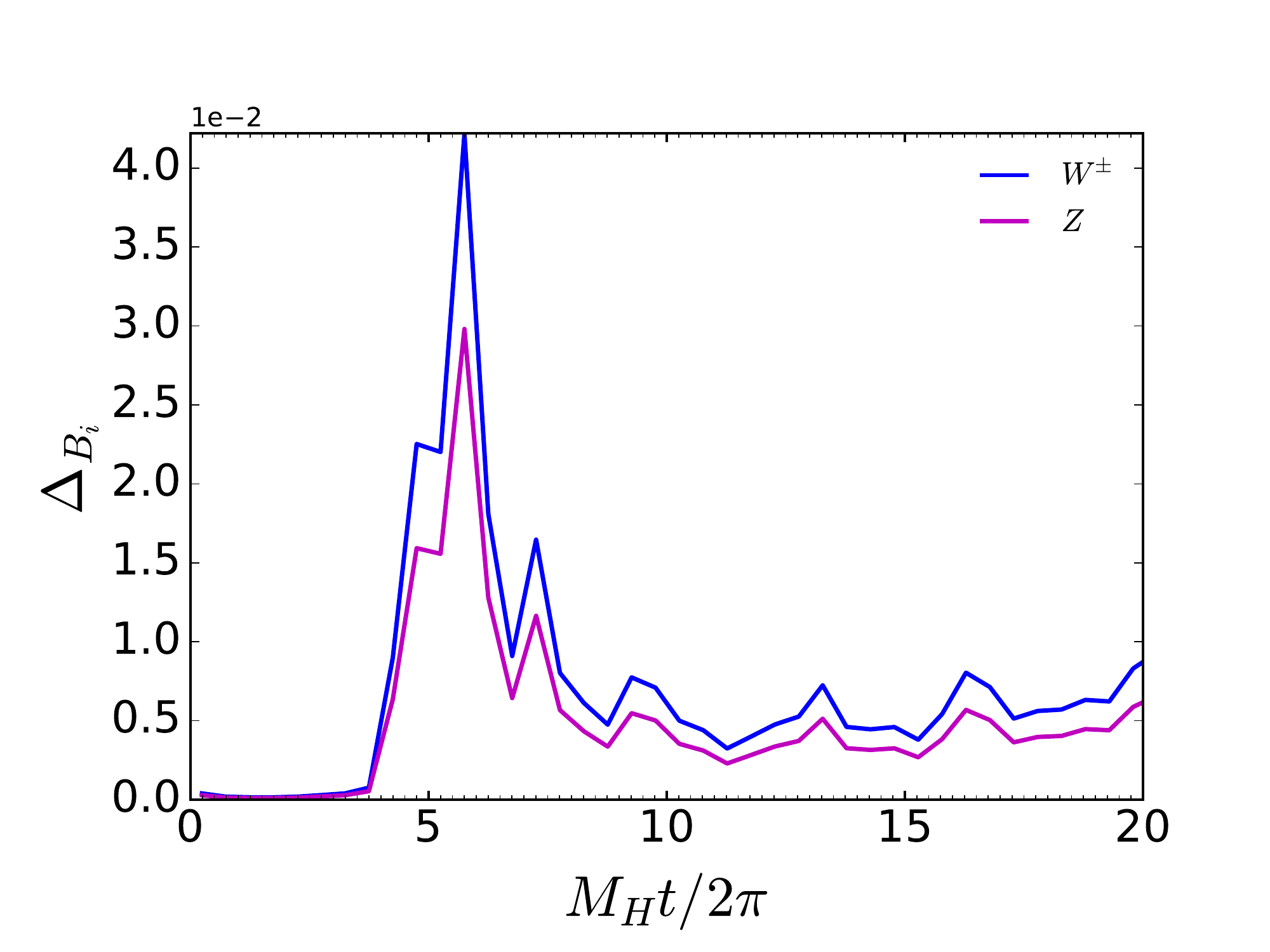}}
\subfigure{\includegraphics[scale=.37]{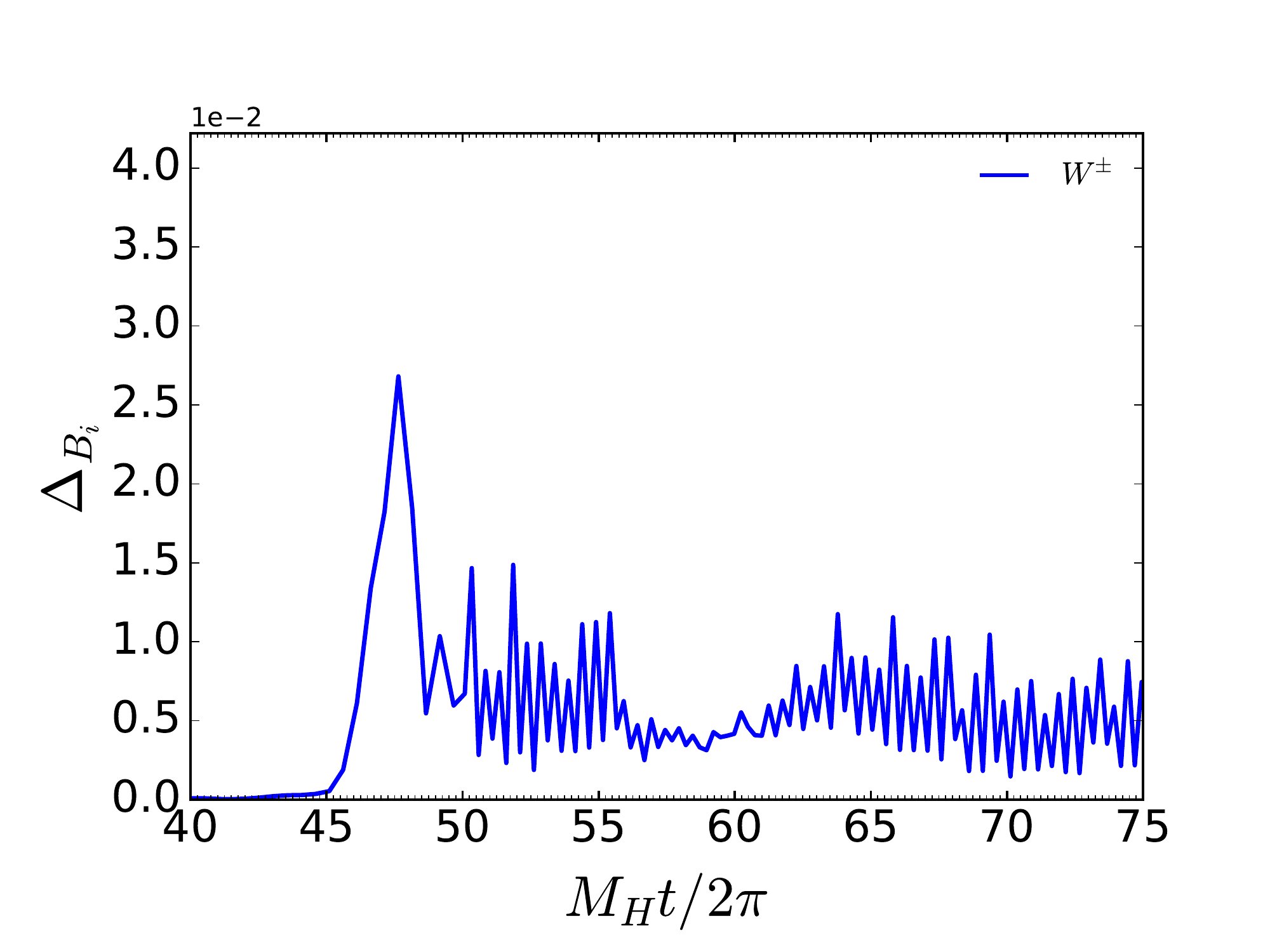}}
\caption{(Left) Evolution of $\Delta_{B_i}$ for $W^{\pm}$ and $Z$ in the absence of fermions. (Right) Evolution of $\Delta_{B_i}$ for $W^{\pm}$ in the presence of fermions.} \label{ratioDelta}
\end{figure}

In spite of the above considerations, non-Abelian interactions could  play an important role once the 
occupation numbers exceed unity. The Hartree approximation \eqref{nonAratio} is not able to capture the effects 
associated to redistribution of momenta. In particular, boson scatterings and annihilations could induce a faster approach to equipartition
by rapidly extending the momentum distribution towards the ultraviolet~\cite{Bodeker:2007fw,Enqvist:2015sua}. Although we 
do not expect order of magnitude corrections to the onset and duration of parametric resonance due 
to this effect, a reliable determination of the condensate depletion time, the approach to equilibrium and the reheating temperature in Higgs inflation requires an analysis that goes far beyond the global modelization of interactions used in this paper.

\item \textit{Gauge boson annihilations}: Once produced, the gauge bosons tend to transfer energy into the SM fermions ($F$) through decays ($B_i\rightarrow F\bar F$) but also through annihilations ($B_i B_i\rightarrow F\bar F $). We disregarded the latest possibility. Although this is certainly a good approximation during the initial stages of \textit{Combined Preheating}, annihilations are expected to be parametrically as efficient as number-changing processes once the gauge boson occupation numbers become large ($ \Gamma_{B_i} \sim \sigma n_{B_i}$)~\cite{Kurkela:2011ti}.

\item \textit{Condensate-Fermion interaction}: We neglected the direct interaction between the Higgs condensate and the SM fermions.\footnote{Remember that direct fermion production is restricted by Pauli-blocking effects. In the analysis of Section \ref{sec2_2} fermions were only produced as a secondary product of the intermediate bosons $B_i$.}Although this can be considered a reasonable approximation from the point of view of fermion production, it is certainly not satisfactory from the point of view of backreaction since  fermions play an important role in the redistribution of momenta \cite{Greene:1998nh,Giudice:1999fb,GarciaBellido:2000dc,Greene:2000ew,Peloso:2000hy,Berges:2010zv}. Higgs-fermion interactions are expected to be relevant when the number of fermions in the plasma becomes large. An estimation of this effect can be obtained by including phenomenological friction terms proportional to the energy density of fermions in the equations of motion for the Higgs field and the gauge bosons. The output of this procedure is presented in Appendix \ref{app3}. These results confirm the naive expectation:  the energy stored into the Higgs
condensate decreases faster when a direct interaction between the fermions and the Higgs condensate is included. Going 
beyond this phenomenological approach would require a proper implementation of fermions 
on the lattice ~\cite{Borsanyi:2008eu,Saffin:2011kc,Saffin:2011kn,Mou:2013kca} or a Boltzmann approach beyond LATTICEEASY. 
\end{itemize}

\section{Conclusions}\label{sec_conc}
We used classical lattice simulations in 3+1 dimensions to study the interplay between non-perturbative boson production at the bottom of the inflationary potential and their subsequent decay into a set of secondary species. The general idea was implemented in a toy version of Higgs inflation in which the interactions among the Higgs field and the SM gauge bosons were replaced by scalar interactions with the SM structure.  Our numerical results extend the analytical estimates in the literature beyond the linear regime and are robust to modifications of the lattice parameters. 

Secondary species were shown to play an important role on the depletion of the inflaton field. In the absence of fermions, the decay of the inflaton is far for complete and the total energy of the Universe becomes democratically distributed among the different species. The 
inclusion of fast and inefficient decays translates into a delay of parametric resonance but also into 
a full depletion of the inflaton component.

While we have only considered a simplified version of Higgs inflation, it would be interesting to extend the analysis to the realistic Higgs inflation scenario 
by properly taking into account the fermionic degrees of freedom and the gauge character of the SM interactions. This non-trivial extension would allow to
obtain a reliable estimate of quantities, such as the depletion time of the condensate or the reheating temperature, which play a central role in 
the phenomenological viability of Higgs inflation if the SM vacuum is not completely stable.\footnote{As shown in Ref.~\cite{Bezrukov:2014ipa}, the fate of
the Universe in (non-critical) Higgs inflation strongly depends on the ability of thermal effects to modify the inflationary potential after the end of  inflation (see also Ref.~\cite{Enckell:2016xse}).}

\section*{Acknowledgments}
We are especially grateful to Mikhail Shaposhnikov for numerous and enlightening discussions during the whole 
development of this work. We also thank Igor Tkachev for discussions during 
the first stages of the project. JR thanks Oscar Garcia-Montero for comments on the manuscript. 
This work was partially supported  by the 
Swiss National Science Foundation and the TRR33 ``The Dark Universe" project.
\appendix
\clearpage

\section{Analytical estimates} \label{app2}
This appendix contains a summary of the main results presented in Refs.~\cite{GarciaBellido:2008ab,Bezrukov:2008ut}. Although limited to the first stages of parametric resonance, these analytical estimates provide a valuable insight on the dynamic of the system under consideration. In particular, they give rise to reasonable order of magnitude predictions for the typical particle momenta to be covered in our lattice simulations and the temporal scale of the problem under consideration. 
\subsection{Boson production in the absence of fermions}

In the quadratic potential \eqref{quadraticP}, the Universe expands as in a matter-dominated background ($a\propto t^{2/3}$) with average energy density $\rho_\chi(t)=\frac{1}{2} M^2 \chi(t)^2$ and zero pressure. Just a couple of oscillations after the end of inflation, the evolution of the Higgs field can be well approximated by the oscillatory function
\begin{equation}\label{phievol}
\chi(t)=\frac{\chi_0\sin(\tau )}{\tau}=\frac{\chi_0\sin(\pi j)}{\pi j}\,,
\end{equation}
with $\tau\equiv Mt$, $j$ the number of semioscillations  and $\chi_0=\sqrt{8/3}M_P$ an initial amplitude dictated by the covariant conservation law $\dot \rho_\chi+3 H \rho_\chi=0$. Taking 
into account Eq.~\eqref{phievol} and the standard field redefinition $B_k\rightarrow a^{-3/2} B_k,$\footnote{The redefinition introduces terms proportional to $H^2$ and $\ddot a/a$ that can be safely neglected at scales smaller than the horizon.} we can rewrite the evolution equation for the gauge boson perturbations\footnote{To unclutter the notation of this section we will denote the bosonic fields $B_i$ as $B$.} 
\begin{equation}\label{evolB0}
\ddot B_k+3 H\dot B_k+\left(\frac{k^2}{a^2}+ \tilde m_B^2(t)\right) B_k =0\;, 
\end{equation}
in the region $M_P/\xi<\chi <\sqrt{3/2} M_P$ as
\begin{equation}\label{evolB}
-B_k'' - \frac{q_B}{j}|\tau| B_k = K^2 B_k\;, \hspace{10mm} q_B \equiv \frac{g^2\xi\,}{\pi\lambda}\;,
\end{equation}
with $K\equiv \frac{k}{aM}$ a rescaled momentum and the primes denoting derivatives with respect to a rescaled time $\tau=M t$.
\begin{figure}
\begin{center}
\includegraphics[scale=0.7]{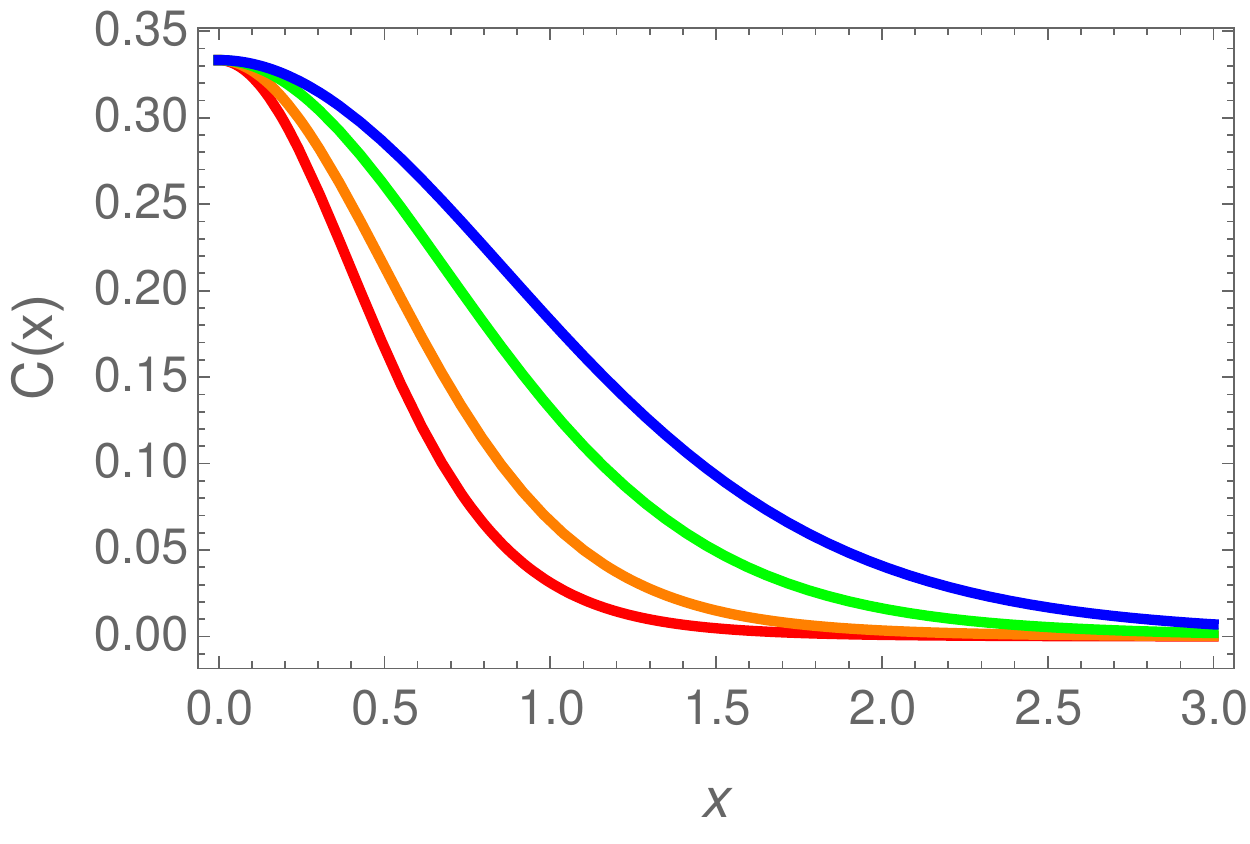}
\end{center}
\caption{The infrared window function $C(x_j)$ for $j = 1, 2, 5$ and $10$ (from left to right). The width of the distribution is mildly depending on the number of semioscillations $j$. For $j=1$, the width is of order  $x_1\sim {\cal O}(1)$, which corresponds to a typical momentum  $k^B_*=q_B^{1/3} M$ with $q_B$ defined in  Eq.~\eqref{evolB}.}\label{CWKB}
\end{figure}
This equation can be interpreted as the Schr\"odinger equation of a particle crossing a (periodic) inverted triangular potential. Using the standard Wentzel-Kramers-Brillouin (WKB) techniques, the number of particles after the $j$-th scattering can be written as a function of the number of particles just before that scattering, $n_{k}(j^-)$, namely
\begin{equation}\label{nk}
n_{k}(j^+) =  C(x_j) + \left(1+2C(x_j) \right)n_{k}(j^-)  +2\cos\theta_{j-1} {\sqrt{C(x_j)\left[C(x_j)+1\right]}}\sqrt{n^2_{k}(j^-)+n_{k}(j^-)}\,, \nonumber
\end{equation}
with
\begin{equation}
C(x_j) \equiv  \pi^2\left[{\rm{Ai}}\left(-x_j^2\right){\rm{Ai}}'\left(-x_j^2\right) + {\rm{Bi}}\left(-x_j^2\right){\rm{Bi}}'\left(-x_j^2\right)\right]^2\,,\hspace{5mm} x_{j} \equiv \frac{K}{(q_B/j)^{1/3}}\,,
\end{equation}
depending on the Airy functions of first and second type and $\{\theta_j\}$ some accumulated phases at each scattering. The momenta distribution of the created particles is dictated by the infrared window function $C(x_j)$. As shown in Fig.~\ref{CWKB}, the typical momentum of the particles created in the first few oscillations is given by\footnote{The scale factor is normalized to the first zero crossing, namely $a(j)=j^{2/3}$.} 
\begin{equation}
x_j\sim {\cal O}(1) \hspace{5mm} \longrightarrow \hspace{5mm}k^B_*(j)\equiv  (q_B^{1/3}M) j^{1/3}\,.
\end{equation}
For the couplings considered in this paper ($\xi = 1500$, $\lambda = 3.4 \times 10^{-3}$, $g_2^2 = 0.3$), this corresponds to an infrared $W$ boson scale $k_*\equiv k^W_*(1)= 34.8\, M$, which we will take as the reference order of magnitude to be covered in our numerical simulations (cf. Section \ref{sec2}). Note also that during the first oscillations, the amplification of perturbations with momenta $k>3k_*$ is completely negligible.  This justifies the choice of the cutoff $k_{\rm \Lambda}$ for the initial fluctuation of the bosonic fields used in our simulations (cf. Section \ref{sec2}).

\begin{figure}\vspace{-0.5cm}
\centering 
\subfigure{\includegraphics[scale=0.37]{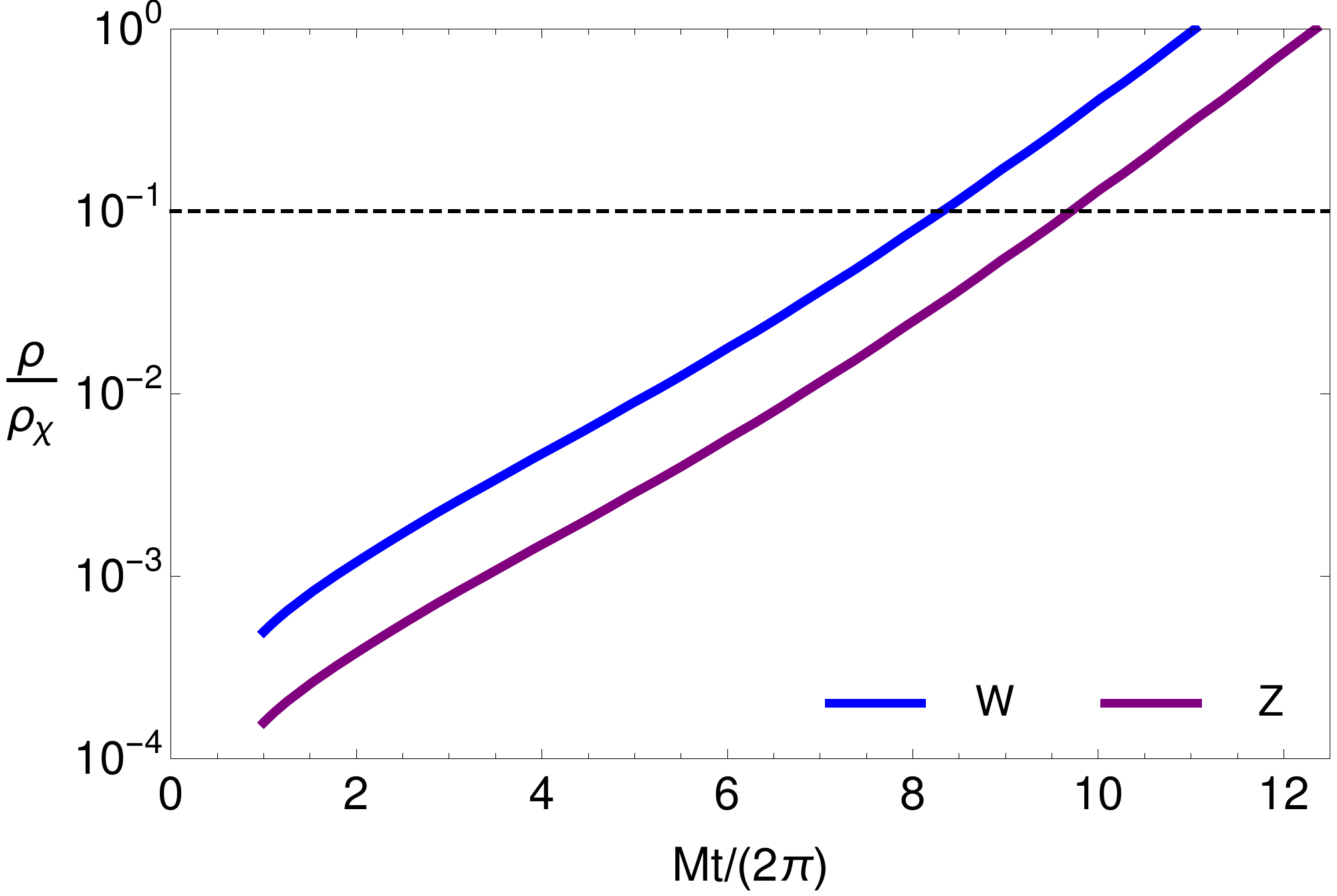}}
\subfigure{\includegraphics[scale=.37]{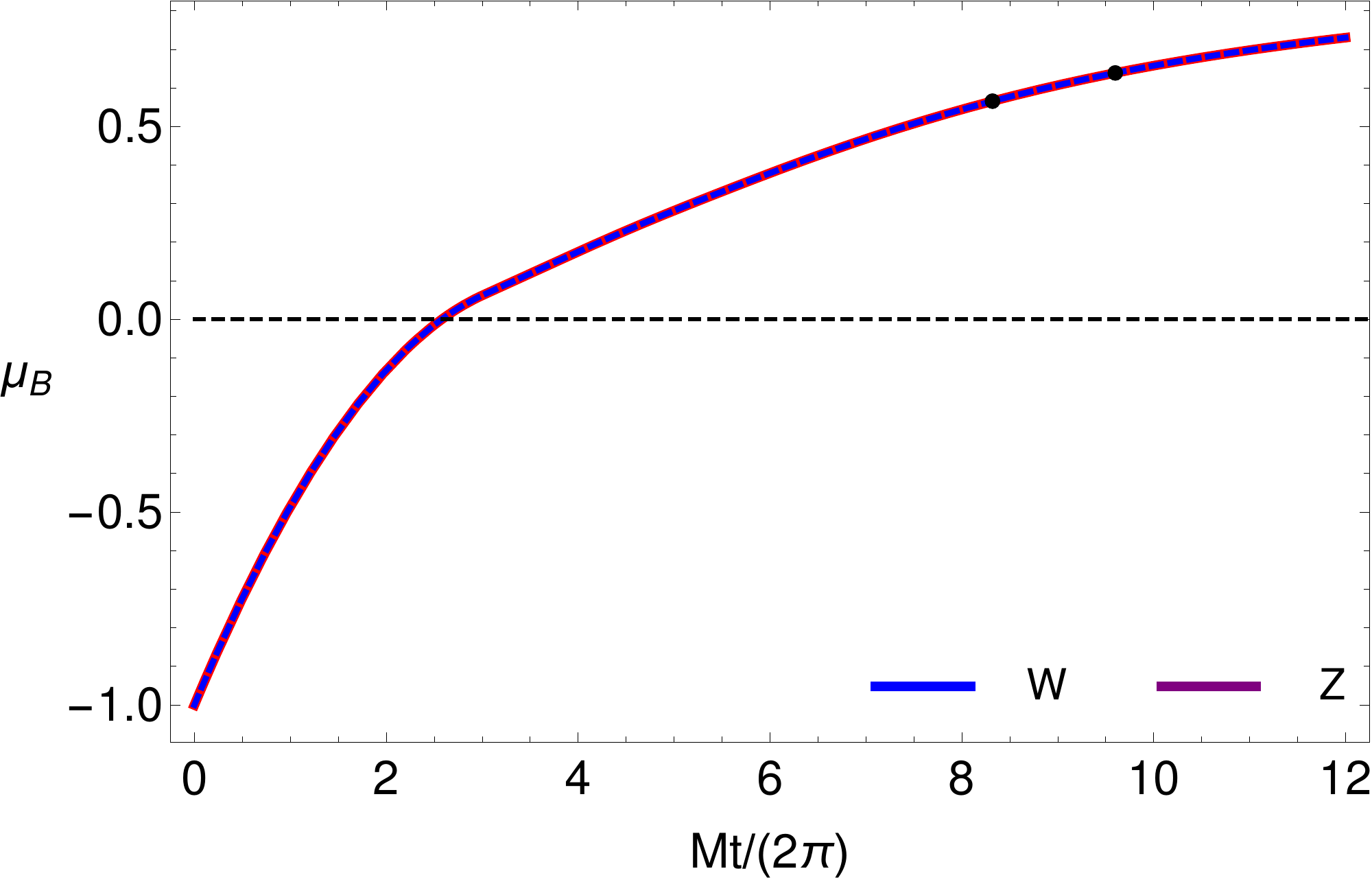}}
\caption{(Left) Analytical estimates for the ratio between the gauge boson energy densities \eqref{Benergy} and the energy density \eqref{Henergy} sitting in the Higgs condensate. (Right) The growing rate \eqref{rateB} for the same species. The black points indicate the values of the $\mu_B$ associated to $ \rho_B/\rho_\chi \simeq  10\,\% $.} \label{analy1}
\end{figure}

As shown in Refs.~\cite{GarciaBellido:2008ab,Bezrukov:2008ut}, the phases  $\{\theta_j\}$ in Eq.~\eqref{nk} can be considered as incoherent  ($\Delta \theta_j  \gg \pi$) for the first few thousands of oscillations. This property allows us to reduce \eqref{nk} to a phase-average relation
\begin{equation}\label{nk2}
\left(\frac{1}{2}+n_{k}(j^+)\right) \simeq \left(1+2\text{C}(x_j)\right)\left(\frac{1}{2}+n_{k}(j^-)\right)\,,
\end{equation}
that can be used recursively to compute the total number of $B$ bosons at each zero crossing
\begin{eqnarray}\label{nB}
\hspace{-12mm}n_B(j^+)=\hspace{-1mm}\frac{1}{2\pi^2 a_j^3}\int dk\,k^2 n_k(j^+) = \frac{q_B M^3}{2\pi^2 j^2} \hspace{-1mm} \int \hspace{-1mm}  dx_1 \,x_1^2 \,n_k(x_{1}\,j^{-1/3}) \,. 
\end{eqnarray}
Eqs.~\eqref{nk2} and \eqref{nB}, together with the average value of the $B$ boson mass between two consecutive crossings
\begin{equation} 
\langle \tilde m_{B} \rangle_j=\frac{\sqrt{3 \pi q_{B}}}{2}  F(j)M\,,
\end{equation}
 \begin{equation}
\label{Fj}
F(j) \equiv\Big\langle \left(1-e^{-\sqrt{2/3}\kappa|\chi(x_j)|}\right)^{1/2} \Big\rangle_j = \frac{1}{\pi}\int_{0}^{\pi} dx_j\left(1-e^{-\sqrt{2/3}\kappa|\chi(x_j)|}\right)^{1/2} \nonumber \simeq \frac{1}{0.57 + 1.94\sqrt{j}}\,,
\end{equation} 
allows to estimate the total energy density of $B$ bosons in the absence of fermions
\begin{equation}\label{Benergy}
\rho_{B}(j) = \langle m_{B} \rangle_{j}n_{B}(j^+)\,.
\end{equation}
The result, normalized to the energy density sitting in the Higgs condensate 
\begin{equation}\label{Henergy}
\rho_\chi=\frac{1}{2}\frac{M^2\chi_0^2}{\pi^2(j+1/2)^2}=\frac{4\xi^2 M^4}{\lambda\pi^2(j+1/2)^2}\,,
\end{equation}
is shown in the left-hand side of Fig.~\ref{analy1}. In the absence of a depletion mechanism, the transfer of energy from the inflaton to the SM particles is very efficient. Indeed, the energy density into $W^{\pm}$ and $Z$ bosons becomes comparable  to the energy stored in the Higgs condensate  ($ \rho_B/\rho_\chi \simeq  10\,\% $) in less than $10$ oscillations. This should be the temporal scale to be expected for the development of parametric resonance in our numerical simulations.

\noindent The right panel of Fig.~\ref{analy1} displays the changing rate of the corresponding energy densities, namely
\begin{equation}\label{rateB}
\mu_B\equiv 2\pi  \frac{d\log \rho_B}{d\left( Mt\right)}\,.
\end{equation}
Note that, in agreement with Eqs.~\eqref{nk2} and \eqref{nB}, this rate \textit{does not depend} on the type of
particle created at the bottom of the potential. In the absence 
of fermions, $W^{\pm}$ and $Z$ bosons grow exactly at the same rate.

\subsection{Boson production in the presence of fermions} 
The creation of fermions out of gauge bosons can proceed through annihilations ($B B \rightarrow F\bar F $) and/or  decays ($B \rightarrow F\bar F$) being the latter the dominant channel at relatively low number densities
 ($ \Gamma_B \gg \sigma n_B$). Assuming that this assumption holds for the numbers of oscillations we are interested in, the number of particles just before a given zero crossing can be written as
 \begin{equation}\label{nk3}
n_k(j^-) =n_k((j-1)^+)e^{-\langle\Gamma_{B_i}\rangle_{j-1}\frac{T}{2}}\;,
\end{equation} 
with
\begin{equation}
\left\langle\Gamma_{B}\right\rangle_j \equiv \frac{2\gamma_{B}}{T} F(j)  \,, \hspace{10mm} \gamma_{B} \equiv  \frac{\pi \alpha_{B}}{2}\left(3 \pi q_{B}\right)^{1/2}\,,
\end{equation}
the average gauge decay width \eqref{decayWZ} between two consecutive crossings.  
Combining Eqs.~\eqref{nk2} and \eqref{nk3}, we obtain the phase-average relation
\begin{eqnarray}\label{iternk}
\hspace{-11mm}\left(\frac{1}{2}+n_k((j+1)^+)\right)\hspace{-1mm}=\hspace{-1mm}A(x_j)\left(\frac{1}{2}+n_k(j^+)\,e^{-\gamma F(j)}\right)\,.
\end{eqnarray}
The iteration of this equation allows us to compute the total number of $B$ bosons at each zero crossing 
\begin{eqnarray}\label{nB3}
\hspace{-12mm}n_B(j^+)=\hspace{-1mm}\frac{1}{2\pi^2 a_j^3}\int dk\,k^2 n_k(j^+) = \frac{q_B M^3}{2\pi^2 j^2} \hspace{-1mm} \int \hspace{-1mm}  dx_1 \,x_1^2 \,n_k(x_{1}\,j^{-1/3}) 
\end{eqnarray}
and the associated energy densities $\rho_{B}(j) = \langle m_{B} \rangle_{j}n_{B}(j^+)$ in the presence of secondary decays. 
\noindent The number of fermions produced between two consecutive scatterings and their corresponding energy density are given by\footnote{The factor 2 accounts for the fact that each gauge boson decays into two fermions.}
  \begin{equation}
  \Delta n_{F}(j)\equiv 2\times \sum_{B} n_B(j^+)(1-e^{-\gamma_BF(j)})  , \hspace{10mm}
\Delta\rho_F{(j)} =\sum_{B} \Delta n_{F}^{(\rm B)}(j) E_{F}^{({\rm B_i})}(j)  \,,
\end{equation}
with
\begin{equation}
E_{F}^{({\rm B})}{(j)}\approx \frac{1}{2}\langle \tilde m_{B} \rangle_j=\frac{\sqrt{3 \pi q_{B}}}{4}  F(j)M \,,
\end{equation}
the  mean energy of the decay products. Summing now over the number of oscillations and taking into account the dilution due to the expansion of the Universe, we obtain the total number of fermions and their total energy density after $j$ semioscillations 
 \begin{equation}\label{ntF}
n_F{(j)} = \sum_{i=1}^{j} \left(\frac{i}{j}\right)^{2}\Delta n_{F}(i)\,,\hspace{10mm}
\rho_F{(j)} = \sum_{i=1}^{j} 
\left(\frac{i}{j}\right)^{8/3}\Delta\rho_F{(i)}\,.
\end{equation}  
\begin{figure}\vspace{-0.5cm}
\centering 
\subfigure{\includegraphics[scale=0.37]{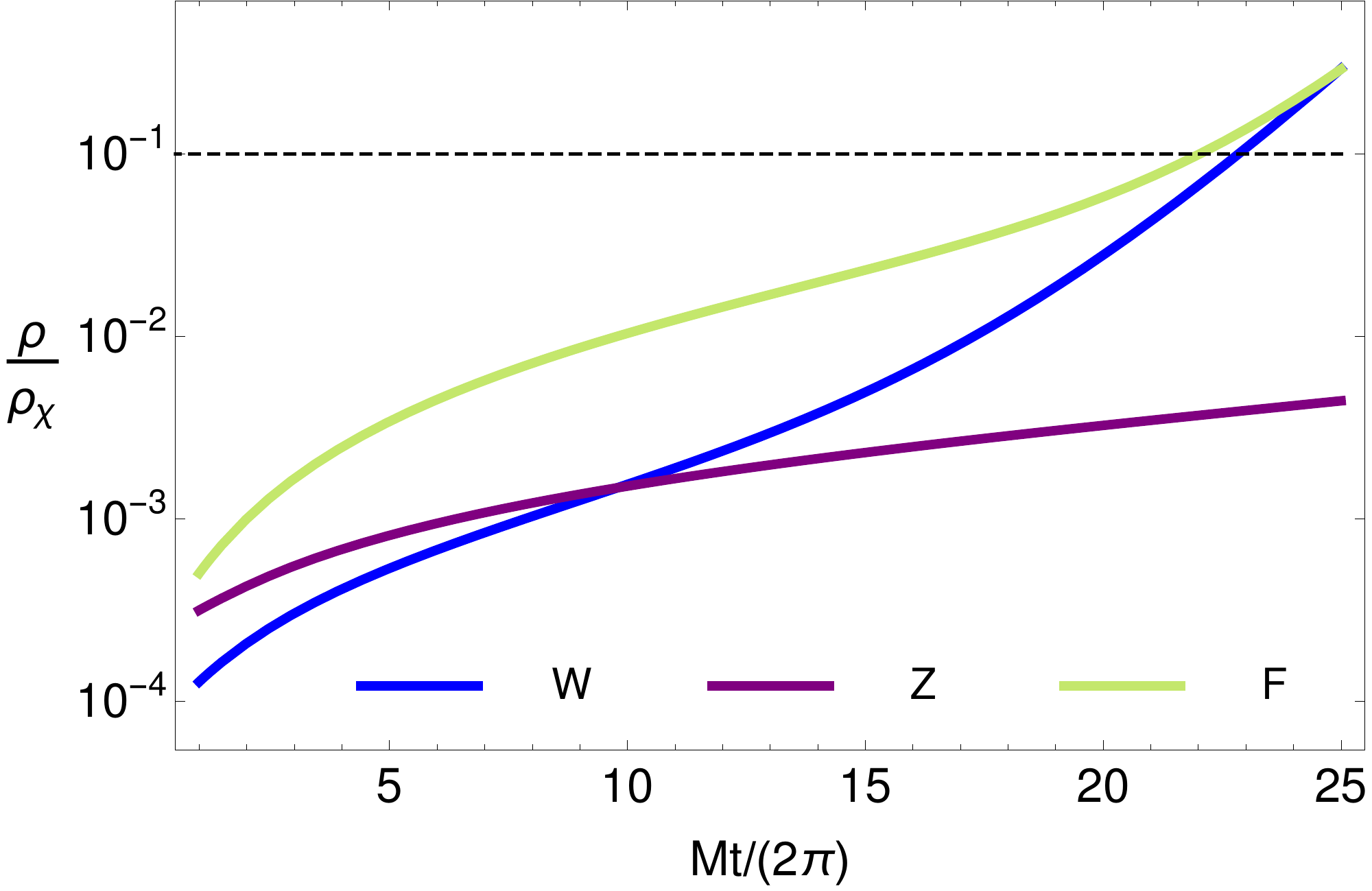}}
\subfigure{\includegraphics[scale=.37]{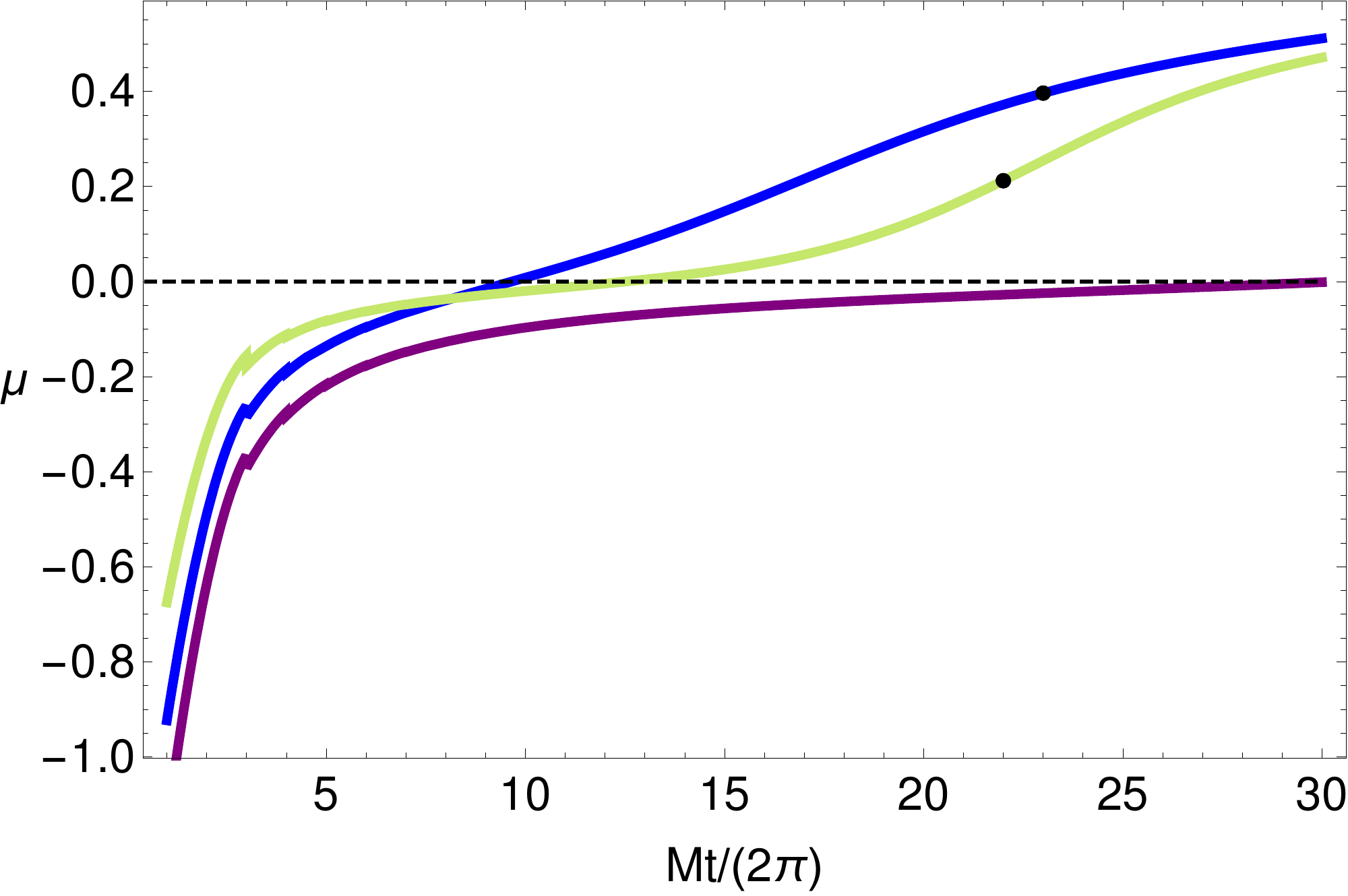}}
\caption{(Left) Analytical estimates for the ratio between the $W^{\pm}$, $Z$ and fermions energy densities and the energy \eqref{Henergy} sitting in the Higgs condensate. (Right) The growing rates \eqref{rateF} for the same species. The black points indicate the values of the $\mu_W$ and $\mu_F$ associated to $ \rho_B/\rho_\chi \simeq  10\,\% $ and $ \rho_F/\rho_\chi \simeq  10\,\% $.} \label{analy2}
\end{figure}
The evolution of the bosonic  and fermionic energy densities computed using the above formalism is shown in the left-hand side of Fig.~\ref{analy2}. As before, we decided to normalize all quantities to the instantaneous energy density of the Higgs condensate, see Eq.~\eqref{Henergy}. The right panel  of Fig.~\ref{analy2} displays the rates
\begin{equation}\label{rateF}
\mu_B\equiv2\pi  \frac{d\log \rho_B}{d\left( Mt\right)}\,,\hspace{10mm}\mu_F\equiv2\pi  \frac{d\log \rho_F}{d\left( Mt\right)}\,.
\end{equation}
The gauge boson decay into fermions is initially very efficient. This translates into an important depletion of the particles created at the bottom of the potential and delays the onset of 
parametric resonance.  Contrary to what happens in the absence of fermions, the gauge boson rate $\mu_B$ \textit{depends} on the particle under consideration, see Eq.~\eqref{iternk}. The
relation $\alpha_Z>\alpha_{W^{\pm}}$ translates into a larger $Z$ boson decay rate ($\gamma_Z>\gamma_W$) and a slower development of parametric resonance in the corresponding channel. As
shown in the left-hand side of Fig.~\ref{analy2}, the energy density into $W^{\pm}$ bosons becomes comparable to that stored in the Higgs condensate  ($ \rho_B/\rho_\chi \simeq  10\,\% $) in
about $22$ oscillations. When this happens, the $Z$ boson energy density is still subdominant and continue to decrease with time due to the expansion of the Universe.

\section{Higgs-fermion interactions and backreaction}\label{app3}

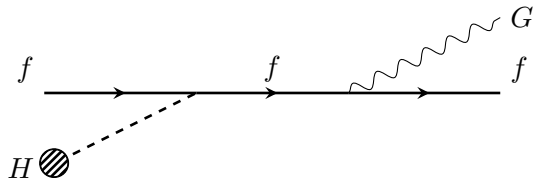
\begin{figure}
\begin{center}
\begin{tikzpicture}[node distance=1cm and 2cm]
\coordinate[label=left:$H$] (e1);
\coordinate[right=of e1] (aux1);
\coordinate[right=of aux1] (aux1b);
\coordinate[right=of aux1b] (e2);
\coordinate[above right=of e1] (aux2);
\coordinate[right=0cm of aux2] (aux0);
\coordinate[right=of aux2] (aux2b);
\coordinate[right=0cm of aux2b] (aux0b);
\coordinate[left=of aux2,label=above left:$f$] (e3);
\coordinate[right=of aux2b,label=above right:$f$] (e4);
\coordinate[above right=of e3] (aux3);
\coordinate[right=of aux3] (aux3b);
\coordinate[left=of aux3] (e5);
\coordinate[right=of aux3b,label=right:$G$] (e6);

\draw[fermionline] (e3) --(aux2);
\draw[fermionline] (aux2) --node[label=above:$f$] {} (aux2b);
\draw[fermionline] (aux2b) --(e4);
\draw[higgsline] (e1) -- (aux0);
\draw[gluon] (aux0b) -- (e6);
		\draw[fill=black] (0.13,0.07) circle (.19cm);
		\draw[fill=white] (0.13,0.07) circle (.18cm);
		\begin{scope}
	    	\clip (0.13,0.07) circle (.19cm);
	    	\foreach \x in {-.9,-.8,...,.3}
				\draw[line width=1 pt] (\x,-.3) -- (\x+.6,.3);
	  	\end{scope}

\end{tikzpicture}
\end{center}
\caption{A typical Higgs condensate-fermion interaction.}\label{Hff}
\end{figure}

The fermions created as secondary products of the intermediate bosons produced at the bottom of the inflationary potential are highly energetic and can play an important role in the redistribution of momenta. The depletion rate of the Higgs condensate due to the rescattering of fermions is expected to be proportional to the number of fermions $n_F$ in the plasma
\be
\frac{dn_s}{dt}= \tilde\Gamma n_s\,, \hspace{10mm} \tilde\Gamma \equiv \sigma n_F\,,
\ee
with $\sigma$ the typical cross-section for a Higgs-fermion interaction as that shown in Fig.~\ref{Hff}.  By dimensional arguments, this cross-section scales as
\be\label{cross}
\sigma\sim \frac{\alpha^2_s}{E^2}\,,
\ee
with $E$ the average energy of the relativistic fermions and $\alpha_s$ the coupling constant. Taking into account that  $E\sim n_F^3\sim \rho_F^{1/4}$, this translates into an effective depletion rate $\tilde \Gamma$ proportional to the energy density of the fermions in the plasma
\be
\tilde\Gamma\sim \rho_F^{1/4}\,.
\ee
Higgs-fermion interactions are therefore expected to become relevant once the energy density sitting into fermions is sufficiently large. A proper modeling of this effect goes much beyond the scope of this paper. In what follows, we will mimic this extra depletion mechanism by introducing additional friction terms in the equations of motion for the Higgs and the gauge boson fields. Since the $W^{\pm}$ bosons evolve in a similar way and the $Z$ boson is not significantly produced in the presence of fermions, we will consider a reduced scenario involving just one gauge field, let's say $W\equiv W^+$. Requiring the conservation of the total energy density  $\dot \rho_T+3 H (\rho_T+p_T)=0$, we obtain the following set of equations
\be
\begin{array}{l}
\ddot\chi + 3H\dot\chi + \alpha\tilde\Gamma\left(\dot\chi- \dot W \right)+V_{,\chi}= \frac{1}{a^2}\nabla^2\chi\,,\\
\ddot W +(3H+\Gamma_{W})\dot W - \alpha \tilde\Gamma\left(\dot\chi- \dot W \right)  + V_{,W} = \frac{1}{a^2}\nabla^2 W\,, \\
\dot\rho_F + 4H\rho_F  = \Gamma_W\dot W^2 + \alpha\tilde\Gamma(\dot W- \dot\chi)^2\,, \label{B4}
\end{array}
\ee
with $\alpha$ a phenomenological parameter encoding the net contribution of all the possible scatterings between the SM quarks and leptons and the Higgs condensate. 

\begin{figure}[h]
	\centerline{\includegraphics[width=0.6\linewidth]{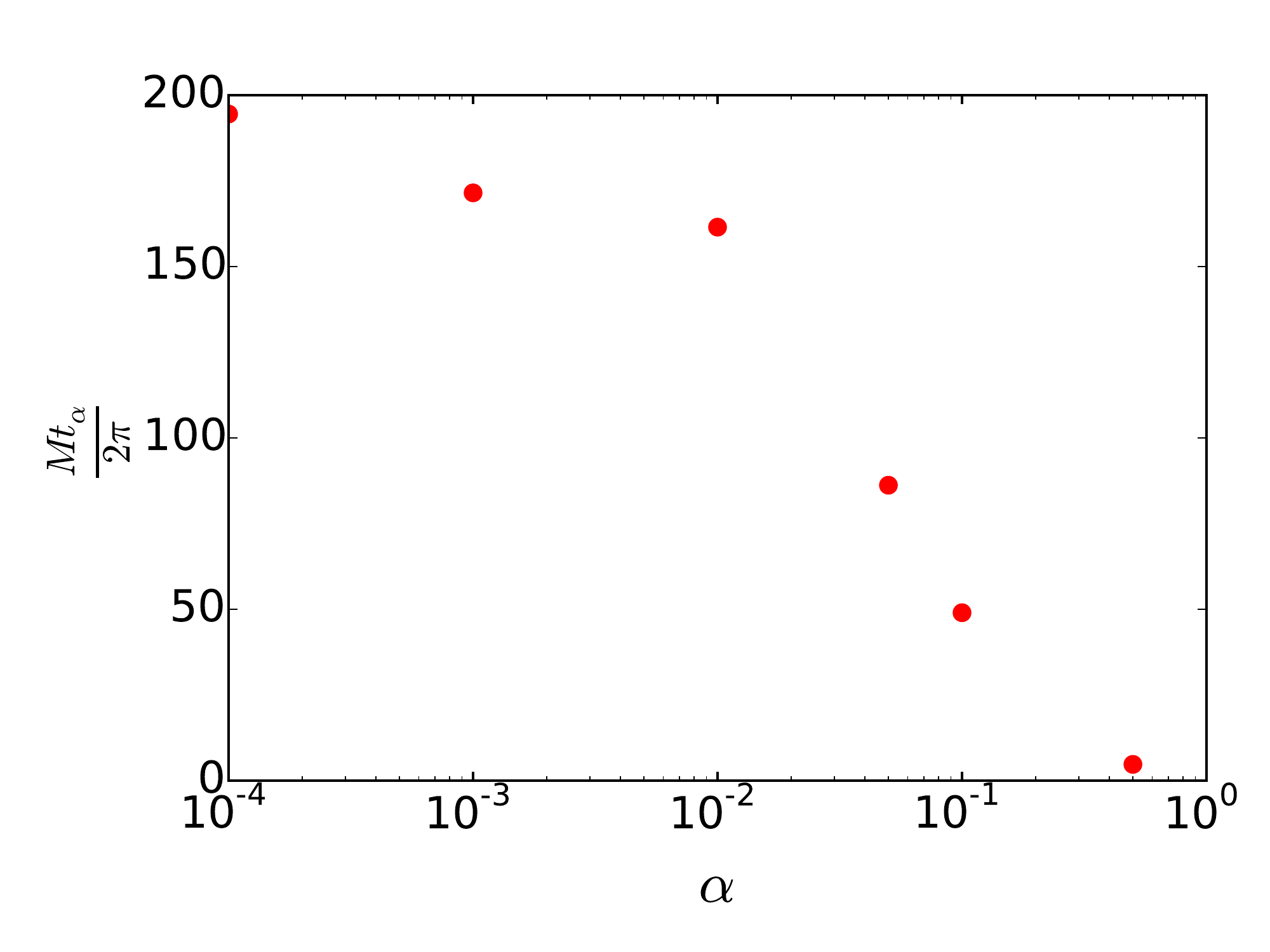}}
	\caption{The time as which the energy of the fermions equals the energy into the Higgs field 
	for different values of $\alpha$.\label{fig:timeOfCrossing}}
\end{figure}

The set of equations \eqref{B4} was solved in a lattice cube of length $L=0.63\, M^{-1}$, $128^3$ points and minimum 
and maximum momentum coverage $k_{\rm min}=0.28\,k_*$ and $k_{\max}=31.61\,k_*$. 
The prototypical effect of Higgs-fermion interactions in the depletion of the condensate is 
summarized in Fig.~\ref{fig:timeOfCrossing}, where we show the number of oscillations $Mt_\alpha/(2\pi)$ at 
which the energy of fermions equals the energy into the Higgs field  for different values 
of $\alpha$. As expected, the energy stored into the Higgs condensate decreases faster when a direct interaction between 
the fermions and the Higgs condensate is included.
 
\newpage

\nocite{*}
\bibliographystyle{plain}

\end{document}